\documentclass[prb,twocolumn,showpacs,preprintnumbers,amsmath,amssymb]{revtex4}
\usepackage{CJK}
\usepackage{graphicx}
\usepackage{amssymb}
\usepackage[latin1]{inputenc}
\usepackage{dcolumn}
\usepackage{bm}
\usepackage{amsmath}
\usepackage{mathtools}
\usepackage{textcomp}
\usepackage{amsbsy}
\usepackage[dvips]{color}
\usepackage{float}
\usepackage{times}
\DeclareGraphicsExtensions{.png,.eps}
\begin{document}

\begin{CJK*}{}{}



\title{RKKY Interactions in Graphene Landau Levels}

\author{Jinlyu Cao, H.A.Fertig, and Shixiong Zhang}
\affiliation{
Department of Physics, Indiana University, Bloomington, IN 47405\\
}

\date{\today}


\begin{abstract}
We study RKKY interactions for magnetic impurities on graphene in situations where the
electronic spectrum is in the form of Landau levels.  Two such situations are considered:
non-uniformly strained graphene, and graphene in a real magnetic field.  RKKY interactions
are enhanced by the lowest Landau level, which is shown to form electron states binding with
the spin impurities and add a strong non-perturbative contribution to pairwise impurity spin
interactions when their separation $R$ no more than the magnetic length.  Beyond this
interactions are found to fall off as $1/R^3$ due to perturbative effects of the negative
energy Landau levels.  Based on these results, we develop simple mean-field theories for
both systems, taking into account the fact that typically the density of states in the
lowest Landau level is much smaller than the density of spin impurities.  For the strain field
case, we find that the system is formally ferrimagnetic, but with very small net moment due
to the relatively low density of impurities binding electrons.  The transition temperature
is nevertheless enhanced by them.  For real fields, the system forms a canted antiferromagnet
if the field is not so strong as to pin the impurity spins along the field.  The possibility
that the system in this latter case supports a Kosterlitz-Thouless transition is discussed.
\end{abstract}
\maketitle
\end{CJK*}
\section{Introduction}
\label{sec:Introduction}
Graphene is one of the most interesting platforms for the two-dimensional electron gas (2DEG) to have become available in the laboratory in recent years,
both for its fundamental physics and for the potential applications it offers
\cite{Castro_Neto_2009,Kotov_2012,Aoki_2014}.  Among its many unique characteristics,
the possibility that it can sustain magnetic order has been an ongoing subject of
investigation. There seems to be little experimental evidence that pristine graphene has such
order \cite{Sepioni_2010}, but theoretical studies strongly suggest that antiferromagnetic order can
be sustained at ribbon edges \cite{Son_2006} or among moments forming on vacancy
defects in the structure \cite{Castro_Neto_2009,Palacios_2010}.  To date there is
no convincing observation of such order, and whether it is realized in the real
material is unclear.  One strategy that has been pursued to enhance magnetism in
graphene is to combine it with magnetic impurities \cite{Dugaev_2006,Brey_2007,Saremi_2007,Black_2010,
Fabritius_2010,Sherafati_2011,Kogan_2011,Lee_2012,
Roslyak_2013,Gorman_2013,Crook_2015,Min_2017}.
In these systems, impurity magnetic moments locally couple to the electron spin density
of the 2DEG, effectively coupling the impurity moments magnetically via RKKY interactions
\cite{RKKY1,RKKY2,RKKY3}.  When the graphene is doped, this leads to
Heisenberg coupling
$J_{RKKY}^{\mu,\nu} \vec{S}_i \cdot \vec{S}_j$ between impurity spins $i,j$,
with the effective exchange constant behaving as
$J_{RKKY}^{\mu,\nu} \sim \sin(k_FR_{ij})/R_{ij}^2$,
where $\mu,\nu$ are the sublattice upon which the impurities at $i$, $j$ reside,
$R_{ij}$ is the impurity separation, and $k_F$ the Fermi wavevector.
In behavior analogous to that of vacancies \cite{Palacios_2010}, the sign
of the coupling changes depending on relative sublattice,
such that $J_{RKKY}^{A,A}=J_{RKKY}^{B,B}=-J_{RKKY}^{A,B}$ \cite{Brey_2007}.

The spatial oscillatory behavior reflects the presence of a Fermi surface, and allows
impurities on the same (opposite) sublattice to couple (anti-)ferromagnetically
up to a separation of order $R_{ij} \sim \pi/k_F$.  By contrast, for undoped graphene,
the point-like form of the Fermi surface \cite{Castro_Neto_2009}
leads to a non-oscillatory form, $J_{RKKY} \sim 1/R_{ij}^3$ \cite{Brey_2007},
which is still opposite for opposite sublattices.  The slow fall-off of this
interaction has interesting consequences for spin stiffness in the system,
for example introducing non-analyticity into the effective energy functional for
spin gradients \cite{Reja_2019}.  In principle this behavior can be
modified by shifting the chemical potential of the system via an electric gate \cite{Reja_2017,Reja_2019}.  The possibility of controlling the magnetic properties
in a single graphene system is one of the reasons it is of such intrinsic interest.

In this paper, we consider alternative strategies to modifying and controlling magnetism
in graphene: application of non-uniform strain \cite{Levy_2010,Jiang_2017,Liu_2018}, or a magnetic field applied perpendicular to the system.
These seemingly different modifications of the graphene system
have in common the restructuring of the electronic spectrum into Landau levels.
While uniform strain has quantitative but not qualitative effects \cite{Gorman_2013} on
RKKY interactions in graphene, non-uniform strain if applied appropriately can
introduce an effective ``pseudo''-magnetic field into the low-energy Hamiltonian
\cite{Guinea_2009,Low_2010}, with the effective field directed oppositely for the
two valleys of the graphene band structure.  Such fields have indeed been created
in graphene bubbles \cite{Lu_2012,Klimov_2012}, and are possible in artificially structured
graphene analogs \cite{Gomes_2012,Tian_2015}.  By its nature, a strain-induced
pseudofield couples to the electronic
orbital degrees of freedom, but to neither the spin of the electrons nor of the
impurities.  A real field, by contrast, couples to spin as well orbital degrees of
freedom, but for low fields this does not fully polarize the impurity spin density,
so that non-trivial order can set in, as we explain below.

The RKKY analysis for these systems differs in important ways from that of graphene in
the absence of a field \cite{Brey_2007}.  The key reason for this is the existence of
a zero-energy Landau level \cite{Gusynin_2005,Goerbig_2011} in which the Fermi level
resides if the system is not strongly doped.  The standard RKKY analysis, which depends
on second order perturbation theory \cite{RKKY1,RKKY2,RKKY3}, becomes invalid because
of the large degeneracy associated with the Landau level.  We demonstrate, however, that
the analytic properties of a Landau level allow one to compute the re-organization of
the energy states in the lowest Landau level due to the presence of two spin
impurities essentially exactly, introducing four bound states that separate off from
the degenerate Landau level, two of which are filled at the electron densities we
consider.  The bound states introduce a spin coupling between the two impurities
which scales {\it linearly} with the $sd$ coupling constant $J$ between an impurity
spin and the 2DEG, and so is formally considerably stronger than standard RKKY
interactions, which scale as $J^2$.  However, the range of this coupling is limited,
falling off as a Gaussian with length scale $\ell=\sqrt{\hbar c/eB} $, with $B$ the
effective field.
Importantly, the remaining Landau levels in the spectrum may be
taken into account perturbatively, yielding an interaction of the same Heisenberg
form $\vec{S}_i \cdot \vec{S}_j$ as in zero field, which is of magnitude $J^2$, but
falls off much more slowly, as $1/R^3$.

An interesting difference between the Landau levels in the strain case
and those of the applied field case is that, in the former, the non-analytic contribution
only applies to the spins on the same sublattice, so that the RKKY couplings
on one sublattice are stronger than on the other.  This raises the
possibility that the system could sustain a {\it net} magnetic moment, so that
the order (at least at the mean-field level) is ferrimagnetic.
In principle this net magnetic moment would make detection of magnetism via magnetization measurements
in these graphene systems considerably easier than the antiferromagnetism
expected of a perfect graphene lattice \cite{Brey_2007}.

To investigate this last effect, we perform a mean-field analysis for strained graphene
that supports Landau levels.  For reasonably size fields, we find that the density of
impurities is actually quite large compared to the density of states in the lowest
Landau levels, so that a model with effective pairwise interactions between
individual impurities becomes inappropriate.  To deal with this situation, our
mean-field theory is developed in terms of two sets of spin impurities, ones that bind
electrons in the lowest Landau level, and ones that do not.  Within this model we
compute the net magnetization as a function of temperature on each of the sublattices,
and find that for reasonable impurity densities and couplings to the electron system,
their difference is quite small, so that direct detection of magnetic order remains
a challenge for these systems.

We also analyze the mean-field phases for the real magnetic field case.  The presence of
Zeeman coupling in both the impurity and electron spins always induces a
magnetization component along the direction of the field.  However, because of
the antiferromagnetic coupling between impurities on opposite sublattices, at weak
applied field the mean-field state is a canted antiferromagnet, with broken U(1)
symmetry, due to spontaneous ordering of the magnetization component of impurity spins
transverse to the applied field.  Formally such long-range order is unstable to
thermal fluctuations at any finite temperature \cite{Mermin_1966}, but the system
can nevertheless support a true thermodynamic phase transition due to vortex excitations
of the U(1) degree of freedom, which support a Kosterlitz-Thouless transition
\cite{Brink_2018}.  This is a distinguishing feature of the graphene - impurity spin system
when it is embedded in a real magnetic field, and we present estimates for the
Kosterlitz-Thouless transition temperature below for reasonable impurity densities
and couplings.

This article is organized as follows.  In Section \ref{sec:model}
we introduce the basic model used for our analysis of RKKY couplings.
Section \ref{sec:RKKYpair} is focused on an analysis of the
coupling strength for a single pair of spin impurities coupled to graphene electrons
with a Landau level spectrum.  In Section \ref{sec:strainMFT} we present our
mean-field theory for the impurity magnetization for the Landau levels
produced by strain, and in Section \ref{sec:fieldMFT} we present the corresponding
analysis for Landau levels produced by a real field.  Finally in Section \ref{sec:sum}
we present a summary and a discussion of implications of and speculations about
our results.

\section{Continuum Hamiltonian}
\label{sec:model}

We begin with a description of the models we adopt to anaylze RKKY interactions
of spin impurities coupled
to graphene Landau levels.  We adopt a continuum model for the Hamiltonian in the vicinity of
a Dirac point at the center of a valley,
\begin{equation}
H^{\tau}_0 = v_F (\tau q_x \sigma_x +  q_y \sigma_y),
\label{H0}
\end{equation}
where $\vec{q}$ is the momentum relative to the $K$ ($\tau=1$) or $K'$ ($\tau=-1$)
points, $v_F$ is the speed of the electrons in their vicinity, and
$\sigma_{x,y}$ are Pauli matrices acting on spinors whose entries encode the
wavefunction amplitude on the $A$ and $B$ sublattices.  In these equations
and what follows, we have set
$\hbar = 1$.  Magnetic
fields, be they effective fields due to strain or a real magnetic field, are
introduced into the orbital Hamiltonian by the Peierl's substitution, $\vec{q}
\rightarrow \vec{q} \pm e \vec{A}$, where $\vec{A}$ is the vector potential,
which for our purposes corresponds to one associated with a uniform magnetic field.
Because $\vec{A}$ is position dependent, the momentum must now be regarded as an
operator, $\vec{q} \rightarrow -i\vec{\nabla}$.  In the case
of a real field, $\vec{q} \rightarrow \vec{q} + e \vec{A}$, and
choosing $\vec{A}=-By\hat{x}$, eigenstates of $H_0^{\tau}$
have the form \cite{Goerbig_2011}
\begin{equation}\label{K_Valley_WF_B_Field}
\psi^{(\tau = + 1)}_{n,k} = \frac{1}{\sqrt{2} } e^{+ i \vec{K} \cdot \vec{r} }
\left(
  \begin{array}{c}
    \phi_{n-1, k}  \\
    \text{sgn}(n) \phi_{n,k}  \\
  \end{array}
\right),
\end{equation}
for $n \ne 0$, and
\begin{equation}\label{K_Valley_WF_B_Field_n=0}
\psi^{(\tau = + 1)}_{n=0,k} = e^{+ i \vec{K} \cdot \vec{r} }
\left(
  \begin{array}{c}
    0\\
   \phi_{n=0, k} \\
  \end{array}
\right)
\end{equation}
for $n=0$, in the $K$ valley.  For the $K'$ valley, the corresponding expressions are
\begin{equation}\label{K'_Valley_WF_B_Field}
\psi^{(\tau = - 1)}_{n,k} = \frac{1}{\sqrt{2} } e^{- i \vec{K} \cdot \vec{r} }
\left(
  \begin{array}{c}
    -\text{sgn}(n) \phi_{n,k}  \\
    \phi_{n-1, k}  \\
  \end{array}
\right),
\end{equation}
for $n\ne 0$, and
\begin{equation}\label{K'_Valley_WF_B_Field_n=0}
\psi^{(\tau = - 1)}_{n=0,k} = e^{- i \vec{K} \cdot \vec{r} }
\left(
  \begin{array}{c}
     \phi_{n=0,k}  \\
    0  \\
  \end{array}
\right)
\end{equation}
for $n=0$.  In these expressions, the wavevector
$\vec{K} \equiv \frac{ 4 \pi}{3 \sqrt{3} a} \hat{x}$ denotes the position of
the $K$ point relative to the $\Gamma$ point in the Brillouin zone, with $a=0.142$ nm the
nearest neighbor carbon bond length, while the
$K'$ point is located at $-\vec{K}$.  The functions $\phi_{n,k}$ are Landau level
states localized near a guiding center coordinate $y_0=k\ell^2$,
$$
\phi_{n,k}= \frac{1}{\sqrt{L_x} }e^{i k x} e^{- \frac{(y-y_0)^2}{2 \ell^2}} N_{|n|} H_{|n|}(\frac{y-y_0}{\ell}),
$$
with $L_x$ the system size in the $\hat{x}$ direction,
and $N_{|n|} \equiv \sqrt{1/{\pi^{1/2}\ell 2^n |n|!}}$ a normalization constant.
$H_{|n|}$ are Hermite polynomials.  The eigenvalues of $H_0^{\tau}$ associated with
these wavefunctions are
$E_{n,k}^{\tau} \equiv E_n^{(0)} =  \text{sgn}(n) v_F \sqrt{2 \hbar e B_s n}.$

For the case of strain-induced magnetic fields, we follow the approach developed in
Ref. \onlinecite{Guinea_2009}.  Briefly, this involves encoding lattice distortions that
vary very slowly on the scale of the graphene lattice constant in a vector potential
given by a strain tensor $u_{ij}$, with
\begin{equation}\label{effective gauge}
  e\vec{A}(\vec{x}) = \frac{\beta}{a}
  \left(
    \begin{array}{c}
      u_{xx}-u_{yy} \\
      -2 u_{xy} \\
    \end{array}
  \right)
\end{equation}
and $\beta = -\frac{\partial \ln t}{\partial \ln a}\approx 2$ specifying the change in
the nearest neighbor tunneling parameter $t$ when the lattice constant $a$ changes.  The
effective gauge field couples with ${\it opposite}$ signs for the two valleys:
$\vec{q} \rightarrow \vec{q} + e \vec{A}$ for the $K$ valley ($\tau=1$), while
$\vec{q} \rightarrow \vec{q} - e \vec{A}$ for the $K'$ valley ($\tau=-1$).
The eigenstates of $H_0^{\tau}$ in this case are
\begin{equation}
\psi^{(\tau = \pm 1)}_{n,k} = \frac{1}{\sqrt{2} }
\left(
  \begin{array}{c}
    \phi_{n-1,\pm k}  \\
    \pm\text{sgn}(n) \phi_{n,\pm k}  \\
  \end{array}
\right),
\label{higher_LLstates_strain}
\end{equation}
for $n \ne 0$, while for $n=0$,
\begin{equation}
\psi^{(\tau = \pm 1)}_{n=0,k} =
\left(
  \begin{array}{c}
    0\\
   \phi_{n=0,\pm k} \\
  \end{array}
\right).
\label{LLLstates_strain}
\end{equation}
Note the important distinction from the real magnetic field case: the support for the two
valleys is essentially the same for each sublattice, whereas in the real field case the roles
of the sublattices is switched.  The difference reflects the fact that the strain breaks the
inversion symmetry of the graphene lattice, so that the two sublattices come in asymmetrically
in the Hamiltonian.  This ultimately opens the possibility that the magnetization can have
different magnitudes on each of the sublattices, with a net magnetization resulting.

To analyze RKKY coupling between impurities carrying spin by the electrons, we will consider
Hamiltonians with two impurities at locations $\vec{R}_{1,2}$ on specified sublattices.
For $\mu_{1,2}=A,B$, the coupling takes the form
\begin{widetext}
\begin{equation}
V^{(\mu_1,\mu_2)} \equiv V^{(\mu_1)} + V^{(\mu_2)}
 = J \left[ \vec{S}_1 \cdot \vec{s} \,\delta(\vec{r}-\vec{R}_1)P_{\mu_1}
 + \vec{S}_2 \cdot \vec{s} \, \delta(\vec{r}-\vec{R}_2)P_{\mu_2} \right].
 \label{sdcoupling}
 \end{equation}
In this expression, $J$ is an assumed $sd$ coupling constant between the impurity
spins and the electron gas, $\vec{S}_{1,2}$ are the impurity spins (assumed classical,
as is standard in RKKY analyses \cite{RKKY1,RKKY2,RKKY3}),
$\vec{s}$ is the electron gas spin operator, and $P_{\mu}$ is a projection operator
onto the $\mu$ sublattice.
Note that the exchange constant $J$ can vary widely depending on the type of impurity
adsorbed on the surface.
For quantitative estimates given below we will adopt a value appropriate for
Co when bound to individual carbon atoms (and so to a particular sublattice),
$J a_C \approx 1$eV, where $a_C$ is the area per carbon atom in the graphene lattice,
with effective spin $S=3/2$ \cite{Fritz_2013}.

For strained graphene, the terms above are sufficient, and the effective Hamiltonian
for a two impurity system is $H_{strain}=\sum_{\tau} H_0^{\tau} +V^{(\mu_1,\mu_2)}$, with
the vector potential properly substituted into $H_0^{\tau}$.  In a real magnetic
field, one must also account for the Zeeman coupling between the field and the
electrons, as well as the impurities.  This introduces terms of the form
\begin{equation}
H_Z = H_Z^{(e)}+H_Z^{(imp)}=g_0\mu_B B s_z+g_{imp}^{(0)}\mu_B \sum_i \hat{z} \cdot \vec{S}_i.
\label{zeemanH}
\end{equation}
Note that $g_0 \approx 2$ for electrons in graphene, and $g_{imp}^{(0)} \approx 2$
for Co adatoms \cite{Fritz_2013}.  For the system in a real magnetic field,
the Hamiltonian for a pair of impurities adsorbed on graphene becomes
$H_{field}=\sum_{\tau} H_0^{\tau} +V^{(\mu_1,\mu_2)} +H_Z$.

We now turn to the computation of the effective RKKY coupling between two impurity
spin degrees of freedom adsorbed on graphene in these situations.

\end{widetext}
\section{RKKY Interaction}
\label{sec:RKKYpair}

As described in the introduction, the computation of RKKY interactions in this problem
necessarily involves a non-perturbative contribution, due to the high degeneracy of a
Landau level.  In particular, if we assume the system to be only moderately doped, so
that the Fermi energy lies in the $n=0$ Landau level states --
which we will from hereon refer to as the lowest Landau level (LLL) -- then one must understand
how these levels become energetically organized in the presence of the impurities.
We begin by showing how this can be done.

\subsection{Lowest Landau Level Energies: Exact Solution}

We begin with the case of a vector potential induced by strain.  From the form of Eq.
\ref{LLLstates_strain}, it is apparent that these states can only couple impurities together
when they are both on the $B$ sublattice, for the specific form of strain we consider,
and we focus for the moment on this case.
With a change to circular gauge, states in the LLL can be written in the
form \cite{Yoshioka_book}
$(0,\phi_{n=0,m}(z))^{\dag}$, where $z=(x-iy)/\ell$, and
\begin{equation}
\label{LLLstates}
 \phi _{n=0,m} (z) \propto z^m e^{-\frac{|z|^2}{4}},
\end{equation}
where $m$ is an angular momentum index. These states have the interesting property
that they are peaked at a distance $r_m = \sqrt{2m}\ell$, and have a width of $\ell$.
A generic state in the
LLL takes the form $(0,f(z)e^{-\frac{|z|^2}{4}})^{\dag}$, with $f$
an analytic function in $z$.  For a finite system, a natural requirement is that
a power law expansion of $f$ contains terms of order smaller than some (large)
integer $M$.  The dimension of the LLL with this condition is $M$.

Now suppose we place the two impurities at positions $\vec{R}_{1,2} = \pm \eta \vec{e}_x$.
Then states of the form
\begin{equation}
\tilde{\phi} _{m} (z) \propto (z^2-\eta^2)z^{m} e^{-\frac{|z|^2}{4}}, m = 0,1,...., M-3
\label{zero_energy_states}
\end{equation}
are completely decoupled from the impurities, and will have the same energy as in their
absence.  It is easy to see that the dimension of this subspace is $M-2$; this means
that the set of states affected by the impurities within the LLL can be reduced to a single
pair, which must be orthogonal to the states in Eq. \ref{zero_energy_states}.  Remarkably,
in the limit $M \rightarrow \infty$ these states may be written explicitly, and take
the form
\begin{equation}\label{xi_orthonormalized}
\begin{aligned}
\xi_{1}(z) = \frac{1}{\sqrt{2\pi \sinh(\frac{\eta ^2}{2})}} e^{-\frac{|z|^2}{4}} \sinh(\frac{\eta z}{2}),  \\
\xi_{2}(z) = \frac{1}{\sqrt{2\pi \cosh(\frac{\eta ^2}{2})}} e^{-\frac{|z|^2}{4}} \cosh(\frac{\eta z}{2}).
\end{aligned}
\end{equation}
Note the states above are orthonormal.

These states as written do not include spin.  When taken into account, we have reduced
the problem of finding the energy spectrum for electrons in the presence
of the two impurities, when projected into the LLL, into a $4 \times 4$ matrix
diagonalization problem.  The energies of these 4 states within the LLL will
be sensitive to the relative orientation of the two spins, and this determines
the LLL contribution to the RKKY interaction between the spins.  Since the LLL states have
zero energy in the absence of the impurities, the Hamiltonian for this case can be taken to
simply be $V^{(B,B)}$ (see Eq. \ref{sdcoupling}).  Writing the single particle states
in the order
$(|\xi_1,\uparrow\rangle,|\xi_1,\downarrow\rangle,|\xi_2,\uparrow\rangle,|\xi_2,\downarrow\rangle)$,
the projected Hamiltonian becomes
\begin{widetext}
\begin{equation}\label{explicit_H}
\begin{aligned}
  \bar{H} &=  \frac{JS}{2\pi}e^{-\frac{\eta^2}{2}}\sqrt{\text{sinh}(\frac{\eta^2}{2})\text{cosh}(\frac{\eta^2}{2})} \;\times\\
  & \; \left\{
\left(
\begin{array}{cccc}
 n^{(1)}_z \sqrt{\tanh (\frac{\eta ^2}{2})} & n^{(1)}_z & \left(n^{(1)}_x-i n^{(1)}_y\right) \sqrt{\tanh (\frac{\eta ^2}{2})} & n^{(1)}_x-i n^{(1)}_y \\
 n^{(1)}_z & \frac{n^{(1)}_z}{\sqrt{\tanh (\frac{\eta ^2}{2})}} & n^{(1)}_x-i n^{(1)}_y & \frac{n^{(1)}_x-i n^{(1)}_y}{\sqrt{\tanh (\frac{\eta ^2}{2})}} \\
 \left(n^{(1)}_x+i n^{(1)}_y\right) \sqrt{\tanh (\frac{\eta ^2}{2})} & n^{(1)}_x+i n^{(1)}_y & -n^{(1)}_z \sqrt{\tanh (\frac{\eta ^2}{2})} & -n^{(1)}_z \\
 n^{(1)}_x+i n^{(1)}_y & \frac{n^{(1)}_x+i n^{(1)}_y}{\sqrt{\tanh (\frac{\eta ^2}{2})}} & -n^{(1)}_z & -\frac{n^{(1)}_z}{\sqrt{\tanh (\frac{\eta ^2}{2})}} \\
\end{array}
\right) \right.\\
&\left. +
\left(
\begin{array}{cccc}
 n^{(2)}_z \sqrt{\tanh (\frac{\eta ^2}{2})} & -n^{(2)}_z & \left(n^{(2)}_x-i n^{(2)}_y\right) \sqrt{\tanh (\frac{\eta ^2}{2})} & i n^{(2)}_y-n^{(2)}_x \\
 -n^{(2)}_z & \frac{n^{(2)}_z}{\sqrt{\tanh (\frac{\eta ^2}{2})}} & i n^{(2)}_y-n^{(2)}_x & \frac{n^{(2)}_x-i n^{(2)}_y}{\sqrt{\tanh (\frac{\eta ^2}{2})}} \\
 \left(n^{(2)}_x+i n^{(2)}_y\right) \sqrt{\tanh (\frac{\eta ^2}{2})} & -n^{(2)}_x-i n^{(2)}_y & -n^{(2)}_z \sqrt{\tanh (\frac{\eta ^2}{2})} & n^{(2)}_z \\
 -n^{(2)}_x-i n^{(2)}_y & \frac{n^{(2)}_x+i n^{(2)}_y}{\sqrt{\tanh (\frac{\eta ^2}{2})}} & n^{(2)}_z & -\frac{n^{(2)}_z}{\sqrt{\tanh (\frac{\eta ^2}{2})}} \\
\end{array}
\right)
\right\}.
\end{aligned}
\end{equation}
In this expression, we have written the two classical spin vectors in the form
$\vec{S}_i=S\vec{n}^{(i)}$ with $i=1,2$.  Notice the result is linearly proportional
to $J$ and $S$; the contribution to the RKKY interaction from these states will be
linear in these quantities, whereas standard RKKY interactions usually contain only
contributions quadratic in these.

If we choose our spin $\hat{z}$ axis to be along
the direction $\hat{n}^{(1)}$, and write
$\hat{n}^{(2)} = (\sin\theta \cos\phi,\sin\theta \sin\phi,\cos\theta)$, the
energy eigenvalues of $\bar{H}$ can be written explicitly in the form
\begin{equation}\label{eigenenergy}
  \begin{aligned}
    \frac{2\pi E_1}{JS} &= -\frac{e^{-\frac{\eta ^2}{2}} \sqrt{\sinh \left(\eta ^2\right)}}{\sqrt{2} \sqrt{\tanh \left(\frac{\eta ^2}{2}\right)}}
\sqrt{\frac{4 (\cos (\theta )+1)}{\left(e^{\eta ^2}+1\right)^2}+2 \sqrt{2} \sqrt{\frac{4 e^{2 \eta ^2} (\cos (\theta )+1)+\cos (2 \theta )-1}{\left(e^{\eta ^2}+1\right)^4}}-\frac{8}{e^{\eta ^2}+1}+4},
\\
    \frac{2\pi E_2}{JS} &= -\frac{e^{-\frac{\eta ^2}{2}} \sqrt{\sinh \left(\eta ^2\right)}}{\sqrt{2} \sqrt{\tanh \left(\frac{\eta ^2}{2}\right)}}
\sqrt{\frac{4 (\cos (\theta )+1)}{\left(e^{\eta ^2}+1\right)^2}-2 \sqrt{2} \sqrt{\frac{4 e^{2 \eta ^2} (\cos (\theta )+1)+\cos (2 \theta )-1}{\left(e^{\eta ^2}+1\right)^4}}-\frac{8}{e^{\eta ^2}+1}+4},
\\
    \frac{2\pi E_3}{JS} &= \frac{e^{-\frac{\eta ^2}{2}} \sqrt{\sinh \left(\eta ^2\right)}}{\sqrt{2} \sqrt{\tanh \left(\frac{\eta ^2}{2}\right)}}
\sqrt{\frac{4 (\cos (\theta )+1)}{\left(e^{\eta ^2}+1\right)^2}-2 \sqrt{2} \sqrt{\frac{4 e^{2 \eta ^2} (\cos (\theta )+1)+\cos (2 \theta )-1}{\left(e^{\eta ^2}+1\right)^4}}-\frac{8}{e^{\eta ^2}+1}+4},
\\
    \frac{2\pi E_4}{JS} &= \frac{e^{-\frac{\eta ^2}{2}} \sqrt{\sinh \left(\eta ^2\right)}}{\sqrt{2} \sqrt{\tanh \left(\frac{\eta ^2}{2}\right)}}
\sqrt{\frac{4 (\cos (\theta )+1)}{\left(e^{\eta ^2}+1\right)^2}+2 \sqrt{2} \sqrt{\frac{4 e^{2 \eta ^2} (\cos (\theta )+1)+\cos (2 \theta )-1}{\left(e^{\eta ^2}+1\right)^4}}-\frac{8}{e^{\eta ^2}+1}+4}.
  \end{aligned}
\end{equation}
\end{widetext}

\begin{figure}
  \centering
  \includegraphics[width=0.3\textwidth]{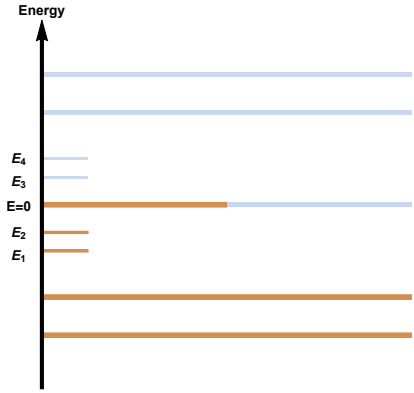}
  \caption{Energy levels from the LLL in the presence of two spin impurities.  Filled(unfilled)
  states are represented in yellow (light blue). Note the presence of both colors at $E=0$,
  indicating the Fermi energy $E_F=0$.}
  \label{2 spins Strain case energy}
\end{figure}

Fig. \ref{2 spins Strain case energy} illustrates the resulting LLL energy structure resulting
from this analysis.  Most states remain at zero energy in spite of the two impurities,
but two break away to negative energy, and two to positive energy.  If the Fermi
energy $E_F$ is in the main band of zero energy states, then the two states at $E_1$ and $E_2$
lower the total electronic energy of the system.  As can be seen from the explicit forms
in Eqs. \ref{eigenenergy}, the amount by which this energy is lowered depends on the
relative orientation of the two spins.

\begin{figure}
  \includegraphics[width=0.9\linewidth]{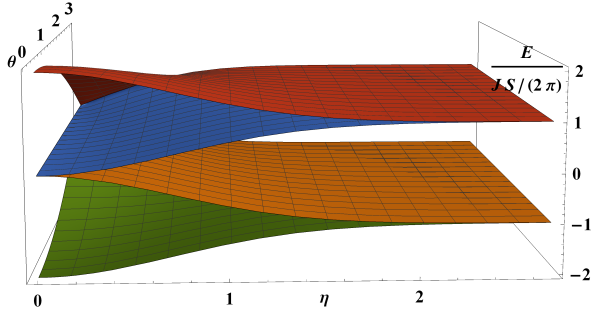}
  \caption{Energies of bound states as a function of separation $\eta$ and relative orientation
  angle $\theta$. }
  \label{fig:3D_E}
\end{figure}

Fig. \ref{fig:3D_E} illustrates the behavior of the four energy levels as a function of
separation $\eta$ and relative orientation angle $\theta$.  Notice that the total lowering
of the energy is always maximized when the two spins are aligned, so that this contribution
to the spin-spin coupling is ferromagnetic.  This is the same sign of coupling for spins
on the same sublattice in the absence of a magnetic field \cite{Brey_2007}.

\begin{figure}
  \centering
  \includegraphics[width=0.5\textwidth]{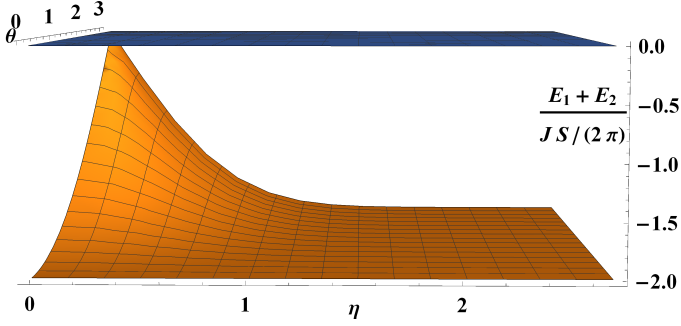}
  \caption{Total LLL energy for electrons in the LLL in a strain-induced magnetic field,
  as a function of relative impurity spin orientation $\theta$ and unitless separation $\eta$.
  Blue plane at the top represents the zero of energy.}
  \label{Total_LLL_Contribution}
\end{figure}

\begin{widetext}

The situation for electrons in a real magnetic field is similar, but one must include the Zeeman
term $H_Z^{(e)}$ (Eq. \ref{zeemanH}) in the Hamiltonian.  In this case the induced interaction
by the LLL is the same for two impurities on the same sublattice, with one
of the two valleys inducing the interaction in each case (see Eqs. \ref{K_Valley_WF_B_Field_n=0}
and \ref{K'_Valley_WF_B_Field_n=0}.) (Spin interactions from the LLL on opposite sublattices
should be negligibly small due the rapidly oscillating relative phase factors in the real space
wavefunctions, $\sim e^{\pm 2i \vec{K} \cdot \vec{r}}$.)  The resulting energies for the
two filled negative energy states for impurities on the same sublattice are
\begin{align}\label{NPT_E1}
  E_1 = & -\frac{1}{2 \pi }
  \left[ \pi  \frac{J S}{\ell^2} g_0 \mu_B B \left((n^{(1)}_z)+(n^{(2)}_z)\right)
  +\frac{J^2 S^2}{\ell^4} \left(\frac{\left(h^2-1\right)^2 X}{\left(h^2+1\right)^2}+1\right)+ \pi ^2 (g_0 \mu_B B)^2 \right.\nonumber \\
        & +\frac{\pi  J S}{(h^2+1)\ell^2}  \left( \frac{2 \left(h^2-1\right)^2 JS g_0 \mu_B B (X+1) \left((n^{(1)}_z)+(n^{(2)}_z)\right)}{\pi \ell^2} \right. \nonumber\\
        &+ \left.\frac{(h^2-1)^2 J^2 S^2 (X+1) \left(h^4 (X+1)-2 h^2 (X-3)+X+1\right)}{\pi ^2 (h^2+1)^2 \ell^4} \right. \nonumber\\
        & \left. \left.  (g_0 \mu_B B)^2 \left(\left(h^2+1\right)^2 \left( (n^{(1)}_z)^2+ (n^{(2)}_z)^2   \right)  +2 \left(h^4-6 h^2+1\right) n^{(2)}_z n^{(1)}_z \right) \right)^{1/2} \right]^{1/2}
\end{align}
and
\begin{align}\label{NPT_E2}
  E_2 = & -\frac{1}{2 \pi } \left[  \pi   \frac{JS}{\ell^2} g_0 \mu_B B \left(n^{(1)}_z+n^{(2)}_z\right)+\frac{J^2 S^2 \left(\left(h^2-1\right)^2 X+\left(h^2+1\right)^2\right)}{\left(h^2+1\right)^2 \ell^4}+ \pi ^2 (g_0 \mu_B B)^2 \right.\nonumber \\
        & -\frac{\pi  J S}{(h^2+1)\ell^2}  \left( \frac{2 \left(h^2-1\right)^2 JS g_0 \mu_B B (X+1) \left(n^{(1)}_z+n^{(2)}_z\right)}{\pi \ell^2} \right. \nonumber\\
        & \left.+\frac{(h^2-1)^2 J^2 S^2 (X+1) \left(h^4 (X+1)-2 h^2 (X-3)+X+1\right)}{\pi ^2 (h^2+1)^2 \ell^4} \right. \nonumber\\
        & \left. \left. +  (g_0 \mu_B B)^2 \left(\left(h^2+1\right)^2 \left( (n^{(1)}_z)^2+ (n^{(2)}_z)^2   \right)  +2 \left(h^4-6 h^2+1\right) n^{(2)}_z n^{(1)}_z\right) \right)^{1/2} \right]^{1/2}.
\end{align}
In these expressions, $X \equiv \cos\theta$ and $h = \sqrt{\tanh (\frac{\eta^2}{2 \ell^2})}$.
These rather complicated expressions can be considerably simplified for most physically
relevant situations, for which the ratio $J/\pi\ell^2 g_0 \mu_B B$ is small; for example,
for our Co estimate we obtain $\sim 0.1$ for this ratio.  (Note this ratio is independent
of the field strength $B$.)  In this case an expansion in this ratio yields for the energy
of the two filled states, to second order,

\begin{align}\label{ELLL}
  E_1+E_2 \approx & \frac{J^2 S^2 \left(2 \left(h^2-1\right)^2 n^{(1)}_z n^{(2)}_z +(h^2+1)^2 (n^{(1)}_z)^2+ (h^2+1)^2 (n^{(2)}_z)^2 - 2  \left( (h^2-1)^2 X+ (h^2+1)^2 \right) \right)}{4 \pi ^2 \ell^4 (h^2+1)^2 g_0 \mu_B  B} \nonumber\\
  &-\frac{JS \left(n^{(1)}_z+n^{(2)}_z\right)}{2 \pi \ell^2 }-g_0 \mu_B  B.
\end{align}

For large separations ($\eta \rightarrow \infty$), the two spins decouple, but the
energy remains dependent on the individual spin orientations:
\begin{equation}\label{ELLL0}
  (E_1+E_2)|_{\eta\rightarrow \infty} = \frac{J^2 S^2 \left((n^{(1)}_z)^2+(n^{(2)}_z)^2-2 \right)}{4 \pi ^2 \ell^4 g_0 \mu_B B}-\frac{JS \left(n^{(1)}_z+n^{(2)}_z\right)}{2 \pi \ell^2}- g_0 \mu_B B.
\end{equation}
Neglecting the constant term, we see that the ${\cal O}(J)$ term effectively
renormalizes the impurity gyromagnetic ratio,
$g_{imp}^{(0)} \rightarrow g_{imp}= g_{imp}^{(0)}+ g_1$,
with $g_1 = -J/\pi \ell^2 g_0 \mu_B B$.  The ${\cal O}(J^2)$ term creates a spin anisotropy
favoring an in-plane spin orientation.  The form for the interaction when the asymptotic
energy ($\eta \rightarrow \infty$) is removed is particularly simple:
\begin{equation}
\label{ELLLasym}
(E_1+E_2) -  (E_1+E_2)|_{\eta\rightarrow \infty} =
-\frac{J^2S^2}{2\pi^2\ell^4 g_0 \mu_B B}e^{-2\eta^2}
\left(n^{(1)}_x n^{(2)}_x +n^{(1)}_y n^{(2)}_y \right).
\end{equation}

Our results for the LLL contribution to the RKKY interaction are summarized in
Fig. \ref{Total_LLL_Contribution}.
For effective magnetic fields generated by strain, a ferromagnetic interaction between
spins on one of the two sublattices is generated.  This interaction scales linearly in $J$
and so is relatively strong for small $J$ at length scales below $\ell$, but falls off rapidly above
this length scale.  At such larger distances, the RKKY interaction becomes dominated by
by the contributions from other Landau levels, as we will discuss in the next subsection.
In the case of real magnetic field, the effect of the electron Zeeman coupling simplifies
the behavior from the LLL, introducing an effective renormalization of impurity spin
$g$-factor at linear order in $J$, and inducing anisotropy in the spin-spin interaction
at quadratic order.  Again, this contribution falls off rapidly for impurity separations
large than $\ell$.

In both cases, for large impurity separations the RKKY interaction is dominated by contributions
from $n \ne 0$ Landau levels.  These can be handled in perturbation theory, as we discuss in the next
subsection.

\end{widetext}

\subsection{$n \ne 0$ Levels: Perturbation Theory}

The underlying physics of RKKY interactions between impurity spins is that the energies
of electrons in the system are modified in a way that depends upon their
relative spin orientation.  In many systems this can be handled at second order in
perturbation theory \cite{RKKY1,RKKY2,RKKY3}.  Because of the degeneracy of
the partially filled $n=0$ Landau level, these states had to be handled carefully,
as discussed above.  The remaining levels can be handled in the
more standard fashion.  For $n<0$, it is possible to show that contributions
to the energy at linear order in $J$ from $V^{(\mu_1,\mu_2)}$ (Eq. \ref{sdcoupling})
vanish from these levels
since they are completely filled, and are thus singlets in the electron spin.   The
first non-vanishing contributions come at second order, with the total change in
energy of these filled levels given by
\begin{equation}\label{delta_E_2_valley_simplify02}
  \Delta E^{(2)} =\sum_{\substack{n<0,\\n' \neq n }} \sum_{k,k'} \sum_{s,s'}
  \sum_{\tau,\tau'}  \frac{ |V_{n'k's'\tau',nks\tau}|^2}{E_n^{(0)}-E_{n'}^{(0)}},
\end{equation}
with $V_{n'k's'\tau',nks\tau}=\langle n'k's'\tau' |V^{(\mu_1,\mu_2)}|nks\tau\rangle$
the matrix element of the perturbation to the electron gas of the two impurity
spins, and $n$,$k$,$s$, and $\tau$ are respectively the Landau level,
wavevector (proportional to guiding center coordinate $y_0$), spin, and valley
indices of a state.  The energies $E_n^{(0)}$ are as given in Sec. \ref{sec:model}.
Note that we do not include the electron Zeeman contribution in the energy
denominator, as this is in practice quite small in comparison with
$|E_n^{(0)}-E_{n'}^{(0)}|$, and so may also be handled perturbatively for the situation
of real magnetic field.  We begin first with the case of Landau levels produced
by non-uniform strain.

\begin{widetext}

\subsubsection{Strain-Induced Landau Levels}

We begin with the case where the impurities are both on the $B$ sublattice.  To compute
$\Delta E^{(2)}$ for this case, which we will call $E^{(2)}_{BB}$, we need to compute
sums over $k$, which in the thermodynamic limit ($L_x \rightarrow \infty$) become
integrals.  The relevant integral has the form
\begin{equation}\label{I_n1n2_integral}
  I_{n_1,n_2} (\vec{R}_1,\vec{R}_2) = \frac{L_x}{2 \pi}\int_{-\infty}^{+\infty} dk \phi_{n_1,k}(\vec{R}_1) \phi^*_{n_2,k}(\vec{R}_2),
\end{equation}
and can be evaluated to yield, for $n_1\leq n_2$,
\begin{equation}\label{I_n1n2_simplified}
  I_{n_1,n_2}(\vec{R}_1,\vec{R}_2) =
  \frac{1}{2 \pi \ell^2} \frac{1}{\sqrt{ 2^{n_1-n_2}(n_2!)/(n_1!) }}
        e^{- \eta^2 + i y_m \Delta x/\ell^2 } (-1)^{n_1-n_2}
        \left( \frac{\Delta y + i \Delta x}{2\ell}\right)^{n_2-n_1} L^{n_2-n_1}_{n_1}(2 \eta^2).
\end{equation}
In this expression, $y_m=(\vec{R}_1+\vec{R}_2)\cdot \hat{y}$,
$\Delta x=(\vec{R}_1-\vec{R}_2) \cdot \hat{x}$,
$\Delta y=(\vec{R}_1-\vec{R}_2) \cdot \hat{y}$, and $L_n^m$ is an associated Laguerre polynomial.
For $n_2<n_1$,
the result is the same , with $n_1 \leftrightarrow n_2$, and $y_1 \leftrightarrow y_2$.
In the case $n_1=n_2$, $I_{n_1,n_2}(\vec{R}_1,\vec{R}_2)$ depends only on
$|\vec{R}_1-\vec{R}_2|=2\eta$, and to simplify the notation we write $I_{n,n}
(\vec{R}_1,\vec{R}_2) =I_n(\eta)$.
Summing over Landau level index,
spin, and valley, after considerable algebra we arrive at the expression
\begin{equation}
E^{(2)}_{{BB}} = -\frac{J^2\vec{S_1}\cdot\vec{S_2}}{4\sqrt{2} \pi^2 \ell^3 v_F \hbar}
\left( \sum_{\substack{n>0,\\n' > 0 }} + 2 \sum_{\substack{n>0,\\n' = 0 }} \right)
    \frac{  I_{n}(\eta) I_{n'}(\eta)
            +  I_{n}(\eta) I_{n'}(\eta) \cos(2\vec{K}\cdot (\vec{R}_1 - \vec{R}_2))}
         {\sqrt{n} +\sqrt{n'}}. \nonumber
\end{equation}
In light of Eq. \ref{I_n1n2_simplified}, this can be cast in the form
\begin{align}\label{RKKY_BB_strain_simplified}
E^{(2)}_{BB} = -\frac{J^2\vec{S_1}\cdot\vec{S_2}}{8\sqrt{2} \pi^2 \ell^3 v_F \hbar}
[1+\cos(2\vec{K}\cdot (\vec{R}_1-\vec{R}_2))] (2 e^{-2\eta^2} )
\left\{ \sum_{\substack{n>0,\\n' > 0 }}
\frac{ L_{n}(2\eta^2) L_{n'}(2\eta^2) }
         {\sqrt{n} +\sqrt{n'}} + 2 \sum_{n>0}
    \frac{ L_{n}(2\eta^2)}{\sqrt{n} }
         \right\},
\end{align}
where $L_n(2\eta^2) \equiv L_n^0(2\eta^2)$ is a Laguerre polynomial.  The Landau index sums
appearing in Eq. \ref{RKKY_BB_strain_simplified} can be computed numerically, and must
be cut off at a maximum value that is determined by the density of electrons in $p_z$
orbitals in graphene, which in turn is the density of carbon atoms.  A similar calculation
for impurities both on $A$ sites yields
\begin{align}\label{RKKY_AA_strain_simplified}
E^{(2)}_{AA} = -\frac{J^2\vec{S_1}\cdot\vec{S_2}}{8\sqrt{2} \pi^2 \ell^3 v_F \hbar}
[1+\cos(2\vec{K}\cdot (\vec{R}_1-\vec{R}_2))] (2 e^{-2\eta^2} )
 \sum_{\substack{n>0,\\n' > 0 }}
\frac{ L_{n-1}(2\eta^2) L_{n'-1}(2\eta^2) }
         {\sqrt{n} +\sqrt{n'}},
\end{align}
while for one impurity on an $A$ site and the other on a $B$ site we obtain
\begin{equation}\label{RKKY_AB_strain_simplified}
  E^{(2)}_{AB}= \frac{J^2\vec{S_1}\cdot\vec{S_2}}{8\sqrt{2} \pi^2 \ell^3 v_F \hbar}
  [1-\cos(2\vec{K}\cdot (\vec{R}_1-\vec{R}_2))](4\eta^2 e^{-2\eta^2})
  \sum_{\substack{n>0,\\n' > 0 }} \frac{L^{1}_{n-1}(2\eta^2) L^{1}_{n'-1}(2\eta^2)}{(\sqrt{n} +\sqrt{n'})\sqrt{nn'}}.
\end{equation}

\begin{figure}
  \centering
  \includegraphics[width=0.7\textwidth]{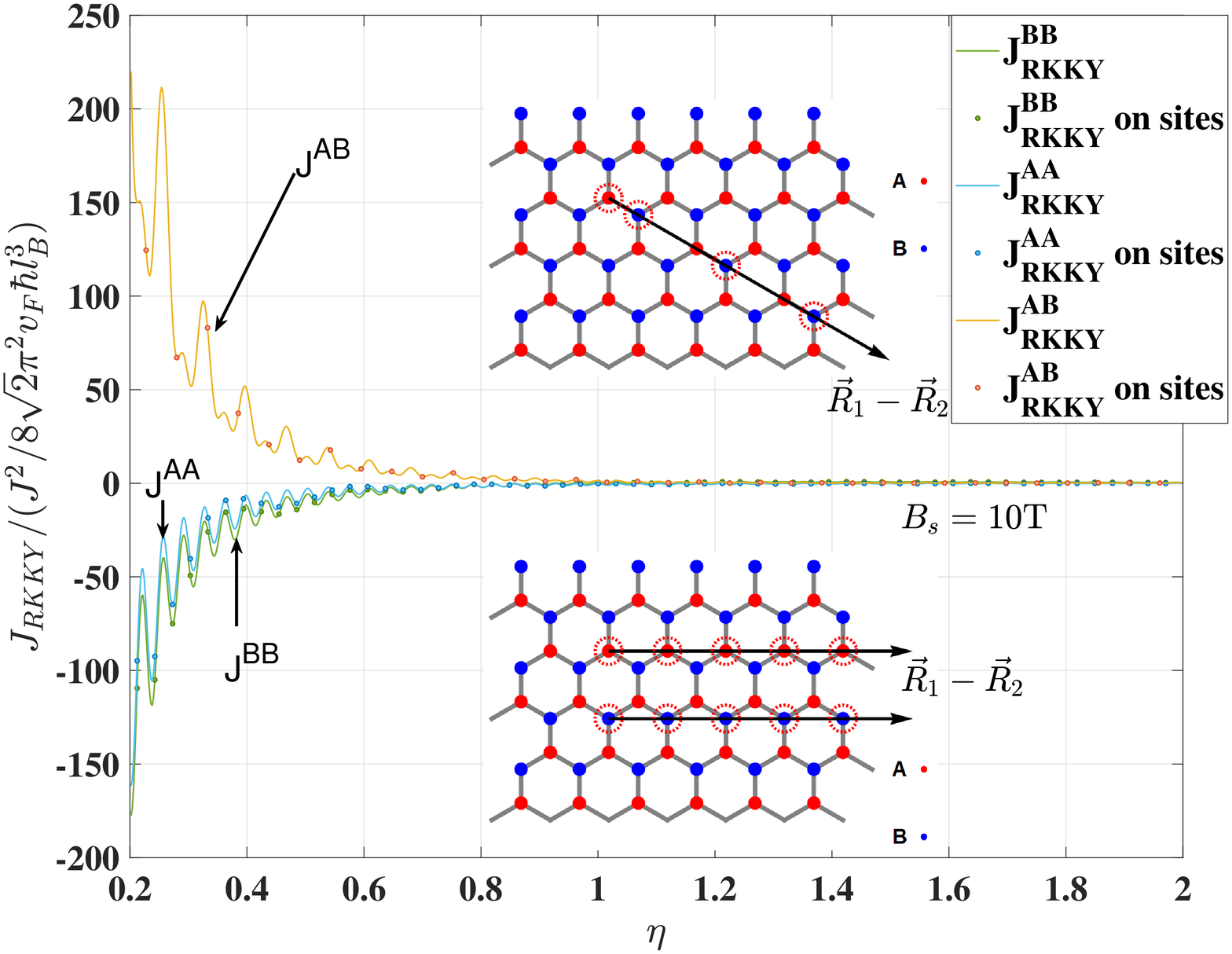}
  \caption{Effective RKKY couplings for impurities on specified sublattices,
  as a function of separation parameter $\eta$, for an effective magnetic
  field of 10T.  Continuous lines show results from Eqs. \ref{RKKY_BB_strain_simplified},
  \ref{RKKY_AA_strain_simplified}, and \ref{RKKY_AB_strain_simplified} for continuous
  $\eta$.  Dots on these lines indicate positions on the honeycomb lattice,
  as depicted in the insets.}
  \label{RKKYstrain1}
\end{figure}

\begin{figure}
  \centering
  \includegraphics[width=0.7\textwidth]{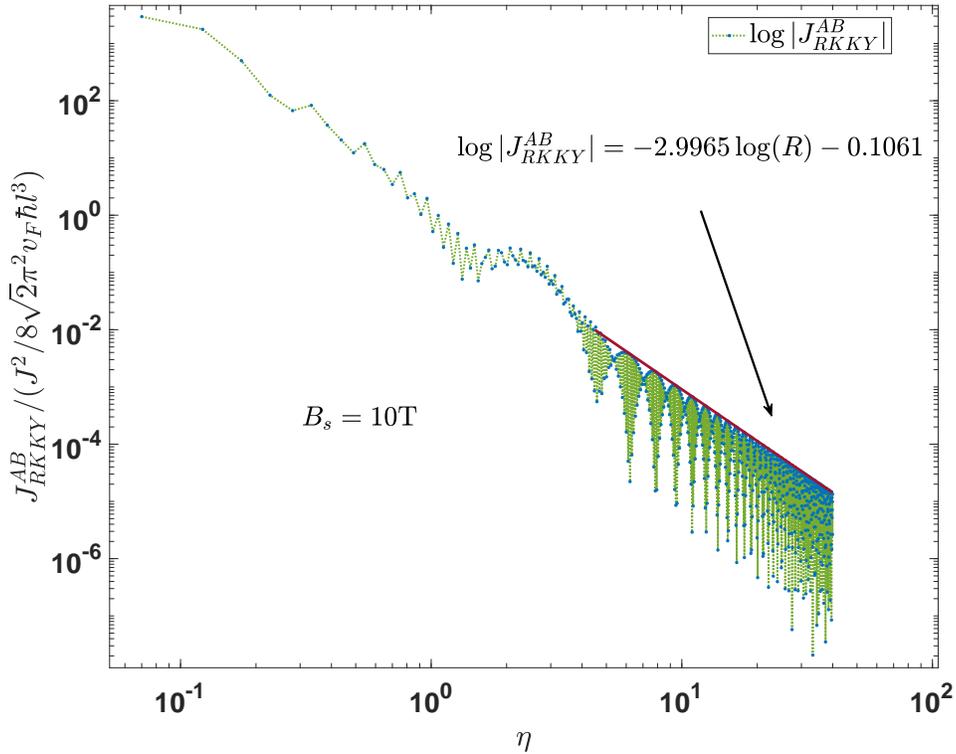}
  \caption{Effective RKKY couplings for impurities on $A$ and $B$ sublattices,
  as a function of separation parameter $\eta$, for an effective magnetic
  field of 10T, illustrating the $1/\eta^3$ falloff of the interaction at
  large separation.}
  \label{RKKYstrain2}
\end{figure}

Writing $E^{(2)}_{\mu\nu} \equiv J_{RKKY}^{\mu\nu}\vec{S}_1 \cdot \vec{S}_2$, we
show representative results for $J_{RKKY}$ in Fig. \ref{RKKYstrain1} as a function of
$\eta=|\vec{R}_1-\vec{R}_2|/2\ell$.  Several points are worth noting.  As in
the case of zero magnetic field, the coupling between spins on the same sublattice
is ferromagnetic, while on opposite ones they are antiferromagnetically coupled.  The presence
of rapid oscillations can be traced back to the need for a cutoff in the sum over
Landau levels, with the largest (negative) Landau index retained determined to
give the correct overall electron density.  
For the most part the
oscillations do not change the sign of the coupling; in the few cases where it does 
this leads neither to a change in
the average sign of the coupling or the sign at very large distances. Finally, the overall scale of
the interaction falls off as $1/\eta^3$, as illustrated in \ref{RKKYstrain2}.

\subsubsection{Magnetic Field-Induced Landau Levels}

The case of Landau level states from a real magnetic field differs from the strain-induced
ones in that there is inversion symmetry, so that the wavefunctions are symmetric
under an interchange of $K$ and $K'$ and the $A$ and $B$ sublattices.  Moreover, the
energies of the electron states in this case are spin-dependent due to the Zeeman coupling.
While the latter effect can in principle be accounted for in $E^{(2)}$ without further approximations beyond the perturbation theory we are using in $V^{(\mu_1,\mu_2)}$,
in practice the Zeeman splitting is very small compared to the Landau level energy
differences without it at any achievable laboratory magnetic field.  Thus it is
sufficient and simplifying to treat the electron Zeeman energy perturbatively as well.

Our perturbation now takes the form
$$
V' \equiv V^{(\mu_1,\mu_2)} + g_0 \mu_B B s_z \equiv V^{(\mu_1,\mu_2)} + V^{(z)},
$$
and, working to linear order in the Zeeman coupling, we use
\begin{align}
|V_{n'k's'\tau',nks\tau}|^2 &\approx |\langle n'k's'\tau' |V^{(\mu_1,\mu_2)}|nks\tau\rangle|^2
+\langle n'k's'\tau' |V^{(\mu_1,\mu_2)}|nks\tau\rangle \langle nks\tau| V^{(z)} |n'k's'\tau'\rangle \nonumber \\
&+\langle n'k's'\tau' |V^{(z)}|nks\tau\rangle \langle nks\tau| V^{(\mu_1,\mu_2)} |n'k's'\tau'\rangle
\label{matrix_element_zeeman}
\end{align}
in the numerator of Eq. \ref{delta_E_2_valley_simplify02}.
(Note that retaining the second order term in $V^{(z)}$ adds a contribution to the energy that
is independent of the relative orientations of $\vec{S}_1$ and $\vec{S}_2$, and so is
irrelevant for our current purpose.)
The computation for the first term of
Eq. \ref{matrix_element_zeeman} runs very similarly to that of the last subsection, and yields the
results
\begin{align}\label{RKKY_BB}
E^{(2)}_{AA}=E^{(2)}_{BB} =& -\frac{J^2\vec{S_1}\cdot\vec{S_2}}{8\sqrt{2} \pi^2 \ell^3 v_F \hbar}
e^{-2\eta^2} \nonumber\\
&\times\left\{ \sum_{\substack{n>0,\\n' > 0 }}
\frac{ L_{n}(2\eta^2) L_{n'}(2\eta^2) + L_{n-1}(2\eta^2) L_{n'-1}(2\eta^2)
            + 2 L_{n-1}(2\eta^2) L_{n'}(2\eta^2) \cos(2\vec{K}\cdot (\vec{r}_1 - \vec{r}_2))}
         {\sqrt{n} +\sqrt{n'}} \right. \nonumber\\
&\left. + 2 \sum_{n>0}
    \frac{ L_{n}(2\eta^2)
            + L_{n-1}(2\eta^2)\cos(2\vec{K}\cdot (\vec{R}_1 - \vec{R}_2))}
         {\sqrt{n} }
         \right\}
\end{align}
and
\begin{equation}\label{RKKY_AB_simplified}
  E^{(2)}_{AB}= \frac{J^2\vec{S_1}\cdot\vec{S_2}}{8\sqrt{2} \pi^2 \ell^3 v_F \hbar}
  [1-\cos(2\vec{K}\cdot (\vec{R}_1 - \vec{R}_2) - 2\Delta \theta)](4\eta^2 e^{-2\eta^2})
  \sum_{\substack{n>0,\\n' > 0 }} \frac{L^{(1)}_{n-1}(2\eta^2) L^{(1)}_{n'-1}(2\eta^2)}{(\sqrt{n} +\sqrt{n'})\sqrt{nn'}},
\end{equation}
where $\Delta \theta$ is the angle between the relative position vector
$\vec{R}_1-\vec{R}_2$ and the $\hat{x}$-direction.  Again writing $E^{(2)}_{\mu\nu} \equiv J_{RKKY}^{\mu\nu}\vec{S}_1 \cdot \vec{S}_2$, Fig. \ref{RKKYfield} illustrates representative results.

\begin{figure}
  \centering
  \includegraphics[width=0.7\textwidth]{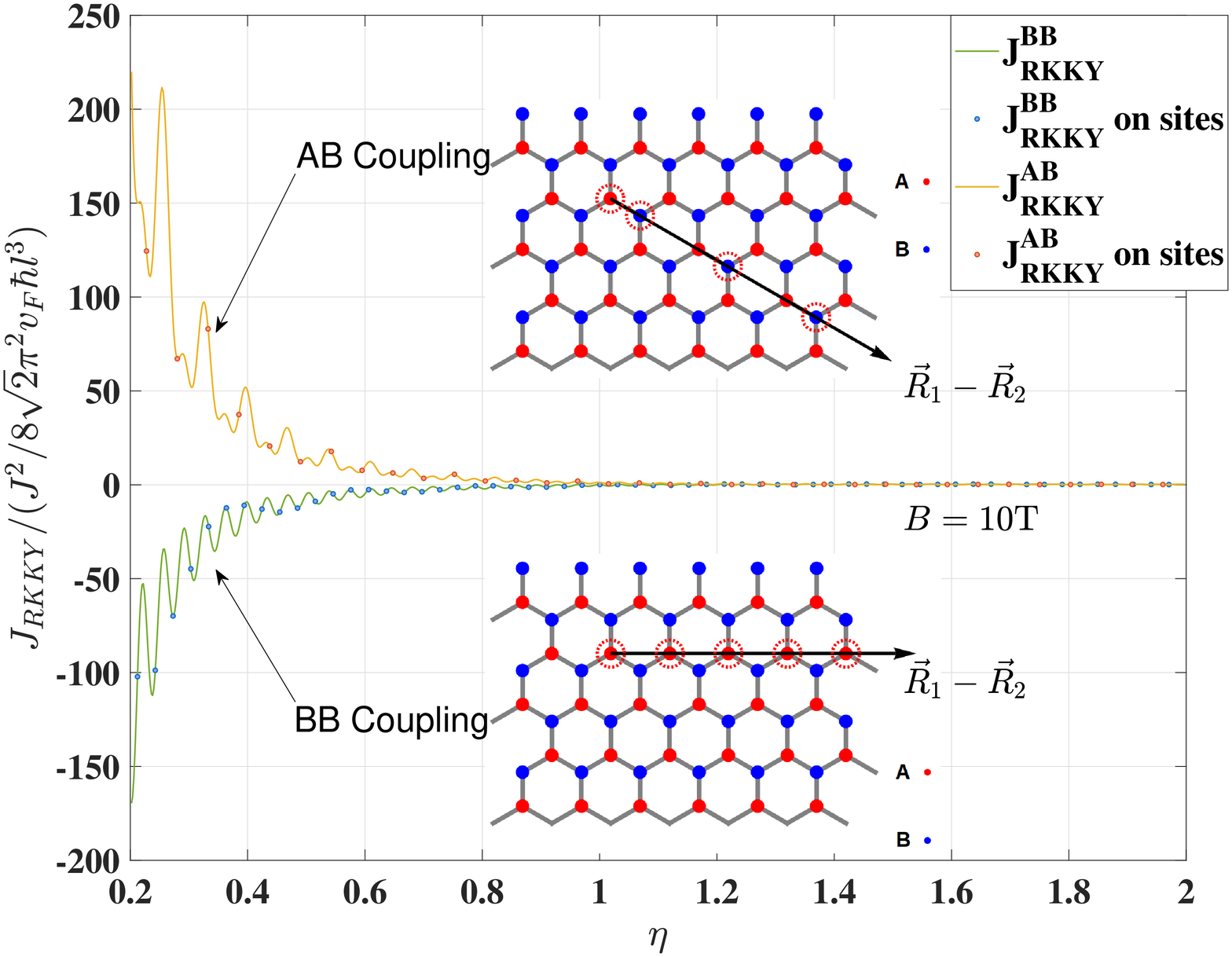}
  \caption{Effective RKKY couplings for impurities on specified sublattices,
  as a function of separation parameter $\eta$, for a real magnetic
  field of 10T.  Continuous lines show results from Eqs. \ref{RKKY_BB},
  and \ref{RKKY_AB_simplified} for continuous
  $\eta$.  Dots on these lines indicate positions on the honeycomb lattice,
  as depicted in the insets.}
  \label{RKKYfield}
\end{figure}

Finally, the electron Zeeman terms yields a contribution to the energy of the form
$$
\Delta E^{(2)}_{z} \equiv \sum_{\substack{n>0,\\n' \geq 0 }} \sum_{k,k'} \sum_{s,s'}
  \sum_{\tau,\tau'}  \frac{ 1}{-\varepsilon_0(\sqrt{n}+\sqrt{n'})}\left(
  V^{\mu_1,\mu_2}_{n'k's'\tau';nks\tau} V^{(z)*}_{n'k's'\tau';nks\tau} +
   +c.c.\right)
$$
with $\varepsilon_0 \equiv \sqrt{2e\hbar B}v_F$. Summing through the discrete indices produces
a result that is independent of which sublattice the impurity spin resides upon, and has the
simple form
\begin{equation}\label{B field impurity indirect couple}
\Delta E^{(2)}_{z} = -\frac{J g_0 \mu_B \vec{B}\cdot(\vec{S_1}+\vec{S_2})}{ 2\pi \ell^2 \varepsilon_0} \sum_{n>0}\frac{1}{\sqrt{n}},
\end{equation}
which is effectively a further renormalization of the impurity $g$-factor.  Note the sum has an
upper cutoff determined by the electron density.

\section{Magnetic Order for Strain-Induced Landau Levels: Mean-Field Theory}
\label{sec:strainMFT}
To assess the effect of the electronic Landau level structure on magnetic order in the
impurity spins, we consider simple mean-field theories of the magnetization.  We begin
in this section with the case of Landau levels induced by strain.  The simplest approach
to this would be treat the system as a system of classical spins, using the
RKKY interactions computed in the previous section as a model for pairwise spin
interactions.  However, in doing this one assumes that the number of states in
the LLL is sufficiently large to provide one bound electron state for every spin
impurity on one of the two sublattices, which we take to be the $B$ sublattice.
In practical situations this turns out not to be the case.  Equating
the density of spin impurities on one sublattice, $n_{imp}$, to the degeneracy of a Landau level
per unit area (including spin), $1/\pi\ell^2$, produces a minimum magnetic field scale
(in Tesla)
$B_c \approx 7.9 \times 10^4 \tilde{n}_{imp}$(T), where $\tilde{n}_{imp}$ is the ratio
of impurity atom density on one of the sublattices
to graphene carbon atom density.  With $\tilde{n}_{imp}$
typically being of order a few percent, we see that the effective field would need
to exceed $\sim 1000$T to reach this limit.  To date,
strain-induced fields \cite{Levy_2010} are at most of order several hundred Tesla,
so that we need to consider situations with fewer LLL states than impurities --
typically, much fewer.

To handle this, we consider a mean field theory in which only a fraction of
impurities, $f_b=1/\pi \ell^2 n_{imp}$, residing on the $B$ sublattice actually form bound states
from the lowest Landau level, while all the spin impurities on both sublattices
interact with one another through the perturbative contributions of the $n \ne 0$
Landau levels.  Denoting the impurity spin degree of freedom at such a bound
site on the $B$ sublattice as $\vec{S}_b$ and a spin degree of freedom on
one of the remaining $B$ sites as $\vec{S}_g$, the total average spin direction for moments
on the $B$ sites becomes
\begin{equation}\label{All B sites average}
  \vec{M}_B  = f_b \langle \hat{S}_b \rangle +(1-f_b) \langle \hat{S}_g \rangle
  \equiv f_b \vec{M}_b +(1-f_b) \vec{M}_g,
\end{equation}
where $\langle \cdot\cdot\cdot \rangle$ here denotes an average over sites and thermal fluctuations.
The corresponding average magnetization (normalized to unity)
for spins on the $A$ sublattice is denoted
by $\vec{M}_A$.  Referring to Eqs. \ref{RKKY_AA_strain_simplified} and
\ref{RKKY_AB_strain_simplified}, we see the pairwise interactions of a spin on an $A$ site
at position $\vec{R}_i$ and either
an $A$ or $B$ site at position $\vec{R}_j$ may be written in the forms
$E^{(2)}_{AA}=J_{RKKY}^{AA}(|\vec{R}_i-\vec{R}_j|)\vec{S}_i \cdot \vec{S}_j$
and $E^{(2)}_{AB}=J_{RKKY}^{AB}(|\vec{R}_i-\vec{R}_j|)\vec{S}_i \cdot \vec{S}_j$,
respectively.  To form a single spin average, we adopt a simple model pair
distribution function
$$
P_{\text{imp}}(R) \propto \tanh(\frac{R}{a}\sqrt{\widetilde{n}_{\text{imp}}})
$$
of finding an impurity on one of the sublattices at a displacement $\vec{R}$
within an area $d^2R$, given that there is an impurity at the origin.  Assuming
the impurity at the origin is on the $A$ sublattice, we can then write an
{\it average} energy functional for its spin of the form
\begin{equation}\label{A site Hamiltonian}
    E_A = \hat{S}_A \cdot \left( \vec{M}_B  \bar{J}_{\text{RKKY}}^{AB}(n_{\text{imp}}) +\vec{M}_A  \bar{J}_{\text{RKKY}}^{AA}(n_{\text{imp}}) \right),
\end{equation}
where $\hat{S}_A$ denotes the orientation of the spin at the origin, and
\begin{equation}
\label{Jbardef}
\bar{J}_{\text{RKKY}}^{\mu\nu}(n_{\text{imp}}) \equiv
S^2\int d^2R_i P_{\text{imp}}(R_i) J_{\text{RKKY}}^{\mu\nu}(R_i).
\end{equation}
Similarly, for a $B$ site lacking bound electrons, the effective energy functional is
\begin{equation}\label{green B Hamiltonian}
    E_g = \hat{S}_g \cdot \left( \vec{M}_B  \bar{J}_{\text{RKKY}}^{BB}(n_{\text{imp}}) +\vec{M}_A  \bar{J}_{\text{RKKY}}^{AB}(n_{\text{imp}}) \right).
\end{equation}
The $\bar{J}_{\text{RKKY}}^{\mu\nu}$ coefficients can be computed numerically using our
results from the previous section and our model $P_{\text{imp}}$.  Results of these
calculation are presented in Table \ref{tab:J RKKY average}.
\begin{table}[]
\begin{tabular}{|l|l|l|l|l|l|}
\hline
$\bar{J}_{\text{RKKY}}^{\mu\nu}$ in eV  ($\times 10^{-5}$)  & 1\% & 2\% & 3\% & 4\% & 5\%\\ \hline
 AA     &    -0.8545  & -2.2997  & -4.0877  & -6.1355  & -8.3965 \\ \hline
 AB     &     1.5404  &  4.0948  &  7.2181  & 10.7650  & 14.6553\\ \hline
 BB     &    -0.8273  & -2.2618  & -4.0441  & -6.0888  & -8.3484 \\ \hline
\end{tabular}
\caption{Numerical values of spatially averaged RKKY coupling strength for a
strain magnetic field of strength $B =10$ T, assuming parameters for Co as
discussed in the Introduction. Top line denotes different impurity concentrations, and AA,AB,BB represent sublattice site locations of the impurities.} \label{tab:J RKKY average}
\end{table}

\end{widetext}

To write an effective energy functional for a spin at a site binding an electron, we need to
revisit the quantum problem yielding the bound state energy.  Recalling in deriving the
RKKY interaction for a single pair of spin impurities, we computed the bound state energies
for two electrons interacting (through the $sd$ Hamiltonian) with classical spins
localized at two sites. For our mean-field estimate,
we will consider the microscopic potential due to a single spin in this collection,
taken to be at the origin, and the {\it average} potential from all other sites
on the $B$ sublattice, which carries average magnetic moment per unit area
$n_{\text{imp}}S\vec{M}_B$.
The potential due to the impurity at the origin has the form
$$
V_{0} \equiv  J S  \hat{S}_b\cdot \hat{\vec{\sigma}} \delta(\vec{r}),
$$
which, when projected into the LLL, couples only to the $m=0$ angular momentum state
(see Eq. \ref{LLLstates}).  Projecting into this one spatial state, the mean-field
Hamiltonian for the electron spin becomes
\begin{equation}\label{H LLL MF}
H_{e} = \frac{J S }{2\pi \ell^2}( \hat{S}_b\cdot \hat{\vec{\sigma}} + \vec{m} \cdot \vec{\sigma} ),
\end{equation}
where we have defined the quantity
\begin{equation}
\label{mdef}
\vec{m} = 2\pi\ell^2n_{\text{imp}}\vec{M}_B.
\end{equation}
Eigenstates of $H_e$ are easily seen to have energy
\begin{equation}\label{Eigen Energy LLL MF}
  \varepsilon_{\pm} = \pm \frac{J S }{2\pi l_B^2} \sqrt{1+m^2+2 \vec{m}\cdot \hat{S}_b}.
\end{equation}
With a bound electron occupying the lower energy state, and adding in the perturbative contribution
to the RKKY interaction, we arrive at a mean-field energy functional for impurity spins
on the $B$ sublattice with bound electrons of the form
\begin{widetext}
\begin{equation}
\label{Eb}
E_b = -g_v \frac{J S }{2\pi \ell^2} \sqrt{1+m^2+2 \vec{m}\cdot \hat{S}_b} +
\hat{S}_b \cdot \left( \vec{M}_B  \bar{J}_{\text{RKKY}}^{BB}(n_{\text{imp}}) +\vec{M}_A  \bar{J}_{\text{RKKY}}^{AB}(n_{\text{imp}}) \right)
\end{equation}
\end{widetext}
where $g_v$=2 is the number of valleys.

Our (classical) mean-field theory now proceeds by computing the average normalized magnetization,
which without loss of generality we can assume to lie in the $\hat{z}$ direction,
$M_{A}=\langle \hat{z} \cdot \hat{S}_A \rangle$, $M_g=\langle \hat{z} \cdot \hat{S}_g \rangle$,
and $M_b=\langle \hat{z} \cdot \hat{S}_b \rangle$, using
Boltzmann
probability distributions at temperature $T$ proportional to
$e^{-E_A/k_BT}$, $e^{-E_g/k_BT}$, and $e^{-E_b/k_BT}$, respectively,
with $M_B$ related to $M_b$ and $M_g$ by Eq. \ref{All B sites average}.
This set of equations can be straightforwardly solved numerically.

\begin{figure}
  \includegraphics[width=\linewidth]{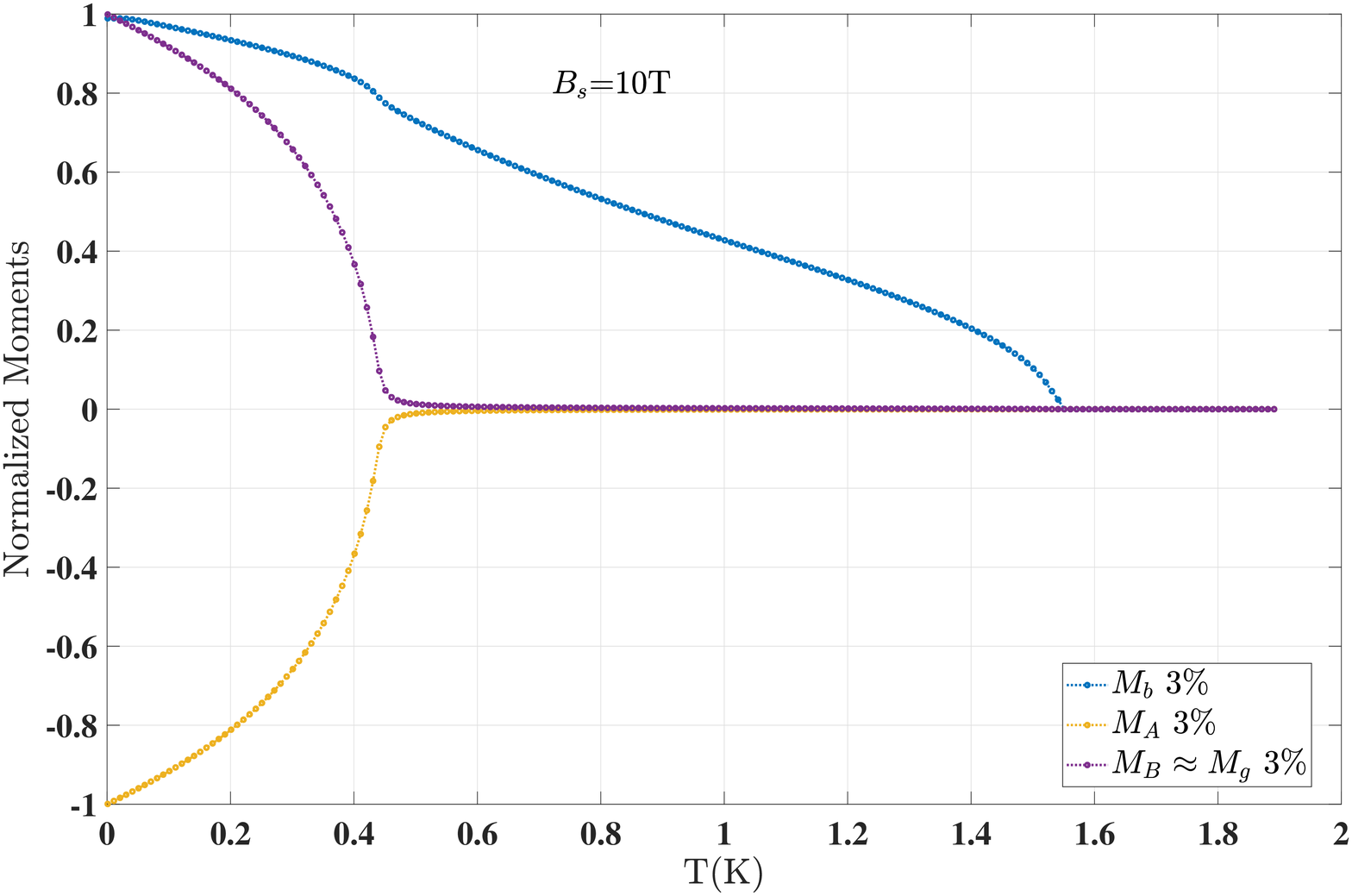}
  \caption{Comparison of $M_A$, $M_B$, and $M_b$ as a function of temperature
  for an effective magnetic field strength of 10T, for parameters relevant to
  Co as described in text.}
  \label{fig:Bseq10}
\end{figure}

\begin{figure}
  \includegraphics[width=\linewidth]{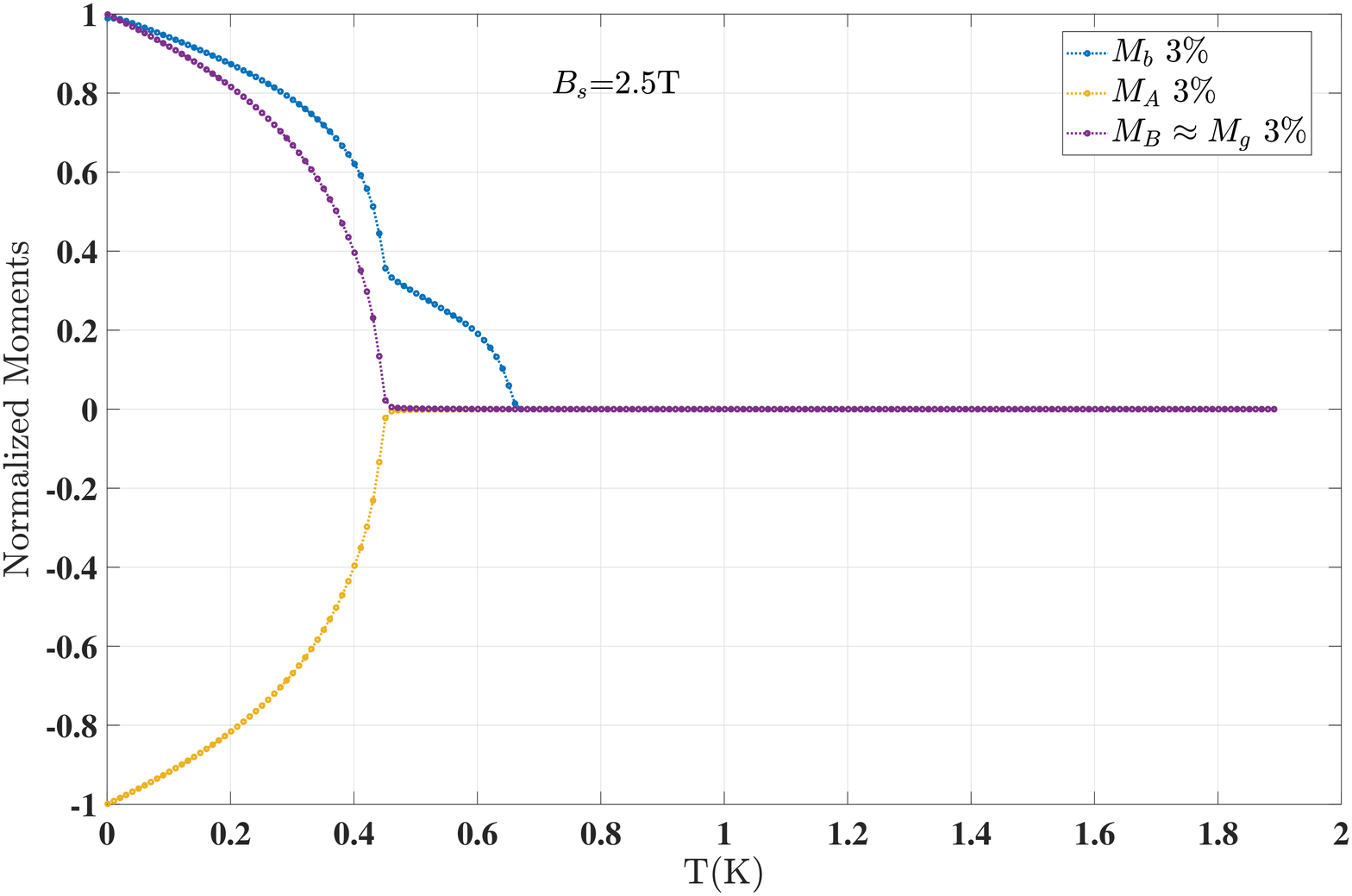}
  \caption{Comparison of $M_A$, $M_B$, and $M_b$ as a function of temperature
  for an effective magnetic field strength of 10T, for parameters relevant to
  Co as described in text.}
  \label{fig:Bseq2.5}
\end{figure}

\begin{figure}
  \centering
  \includegraphics[width=0.4\textwidth]{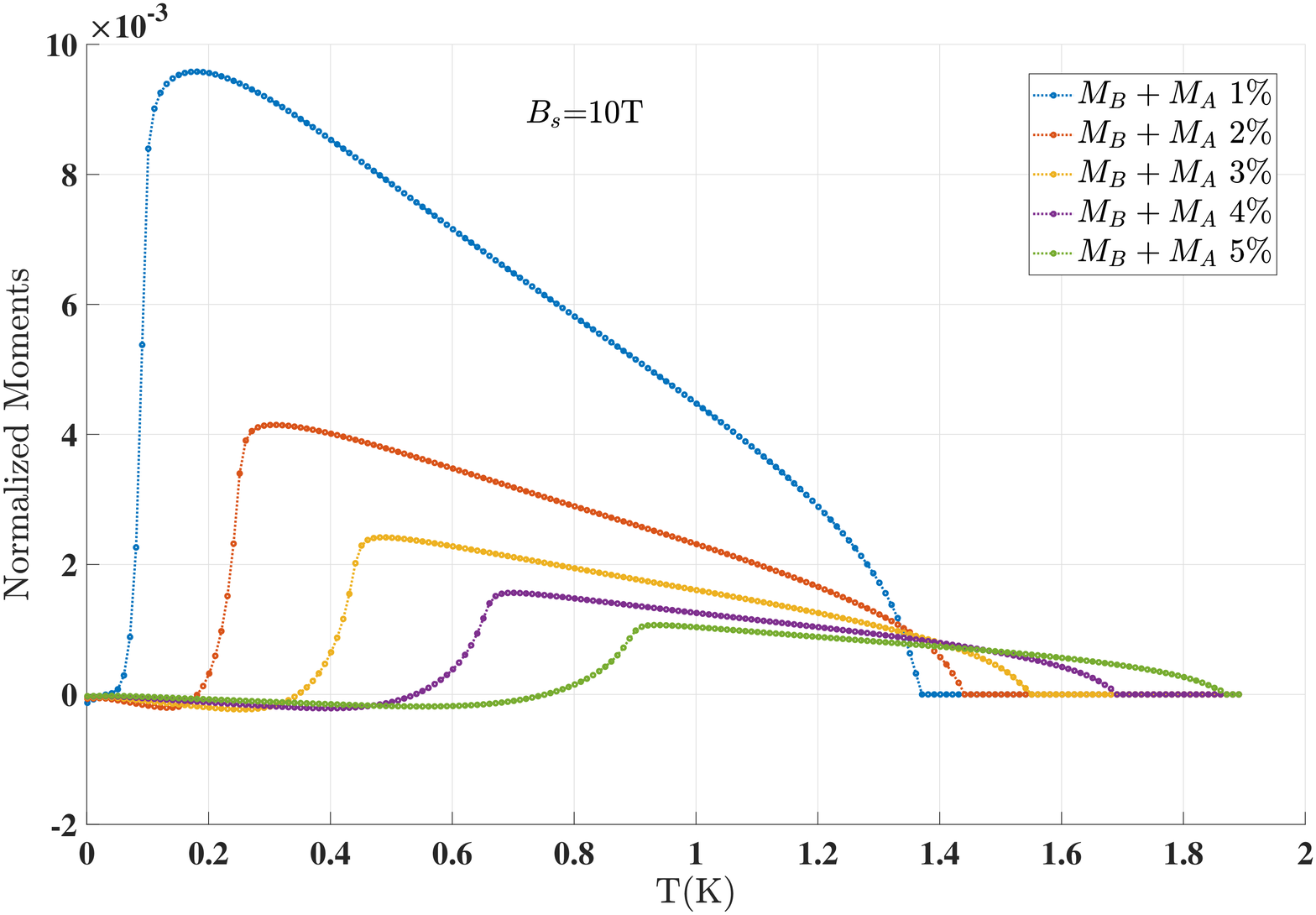}
  \caption{Mean-field net average magnetic moment for strain-induced magnetic field
  as a function of temperature,
  for different coverages $\tilde{n}_{\text{imp}}$ (shown as percent in key). Note this quantity
  always remains small relative to its largest allowed value (1).}
  \label{NetMomFig}
\end{figure}

Figs. \ref{fig:Bseq10} and \ref{fig:Bseq2.5} illustrate typical results for effective fields
of $10T$ and $2.5T$, respectively.  Two features of the curves are of particular note.  First,
$M_A$ and $M_B$ nearly cancel, so that the correlations between sublattices are largely
antiferromagnetic, as is the case for unstrained graphene \cite{Brey_2007}.  However,
$M_b$, the normalized magnetization on the bound electron sites, remains non-zero to larger
temperatures than would be the case without the LLL contribution.  The effect survives to increasingly
high temperature as the field increases, and significantly increases the mean-field
transition temperature.
Unfortunately, because the number of bound electron sites is small,
direct observation of this effect is challenging.  While for any $T>0$, $M_A-M_B \ne 0$,
indicating the system is formally a ferrimagnet, the difference is very small, yielding
only a small net magnetic moment, as illustrated in Fig. \ref{NetMomFig}.
This occurs because $B$ is rather small compared to $B_c$,
so that there is a relatively small number of electrons available in the LLL to bind
to the impurity spins.
An interesting possibility to circumvent this
might be to consider non-uniform strain with small local regions of very high effective
field, where locally an appreciable net magnetization could set in.  Nevertheless, it is
interesting that even in the small (relative to $B_c$) uniform fields we consider,
the enhancement of $T_c$ is considerable.  This may indicate that
the effect of $M_b$ could be seen in thermodynamic measurements which are sensitive
to spin fluctuations.

\section{Magnetic Order for Field-Induced Landau Levels: Mean-Field Theory}
\label{sec:fieldMFT}

We next turn to the formulation of mean-field theory for the case of a real magnetic field.
As discussed in Section \ref{sec:RKKYpair}, the main complication in this situation is
the introduction of Zeeman coupling to the spins, both that of the electrons and those
of the impurity spins.  Moreover, in this more symmetric situation, bound electrons
may appear on either sublattice.  The modifications for the basic energy functionals from the
last section are nevertheless straightforward.
For $A$ and $B$ sites without bound electrons we have
\begin{widetext}
\begin{equation}\label{A site Hamiltonian_zeeman_unbound}
    E_{A,g} = \hat{S}_{A,g} \cdot \left( \vec{M}_B  \bar{J}_{\text{RKKY}}^{AB}(n_{\text{imp}}) +\vec{M}_A  \bar{J}_{\text{RKKY}}^{AA}(n_{\text{imp}}) - S g_{imp}\mu_B \vec{B} \right)
\end{equation}
and
\begin{equation}\label{green B Hamiltonian_zeeman}
    E_{B,g} = \hat{S}_{B,g} \cdot \left( \vec{M}_B  \bar{J}_{\text{RKKY}}^{BB}(n_{\text{imp}}) +\vec{M}_A  \bar{J}_{\text{RKKY}}^{AB}(n_{\text{imp}})- S g_{imp}\mu_B \vec{B} \right)
\end{equation}
respectively.  For an electrons bound to an impurity on sublattice $\mu$
the Hamiltonian (cf. Eq. \ref{H LLL MF}) is modified to
\begin{equation}\label{H LLL MF_zeeman}
H_{\mu,e} = \frac{J S }{2\pi \ell^2}( \hat{S}_{\mu,b}\cdot \hat{\vec{\sigma}} +
\vec{m}_\mu \cdot \vec{\sigma} )-\frac{1}{2}g_0 \mu_B \vec{B} \cdot \vec{\sigma},
\end{equation}
with eigenenergies
\begin{equation}\label{Eigen Energy LLL MF_zeeman}
  \varepsilon_{\pm} = \pm \frac{J S }{2\pi l_B^2} \sqrt{1+h_{\mu}^2+2 \vec{h_{\mu}}\cdot \hat{S}_{\mu,b}},
\end{equation}
in which
\begin{equation}
\label{hdef_zeeman}
\vec{h}_{\mu} \equiv \vec{m}_\mu+\frac{\pi \ell^2g_0\mu_B}{JS} \vec{B},
\end{equation}
and
\begin{equation}\label{mmu}
\vec{m}_\mu = 2\pi\ell^2n_{\text{imp}}\vec{M}_\mu.
\end{equation}
The resulting energy functional for sites
binding an electron takes the form
\begin{equation}
\label{Eb_zeeman}
E_{\mu,b} = -\frac{J S }{2\pi \ell^2} \sqrt{1+h_\mu^2+2 \vec{h}_\mu \cdot \hat{S}_{\mu,b}} +
\hat{S}_{\mu,b} \cdot \left( \vec{M}_\mu  \bar{J}_{\text{RKKY}}^{\mu\mu}(n_{\text{imp}}) +\vec{M}_{\bar{\mu}}  \bar{J}_{\text{RKKY}}^{AB}(n_{\text{imp}}) - S g_{imp}\mu_B \vec{B} \right),
\end{equation}
\end{widetext}
where $\bar{\mu}=B(A)$ if $\mu=A(B)$.
In analogy with the previous section, $\hat{S}_{\mu,b}$ and $\hat{S}_{\mu,g}$ are thermally
averaged with these energy functionals, and we search for self-consistent solutions satisying
\begin{align}\label{zeeman_self_consistency}
\vec{M}_\mu &= f_b\langle \hat{S}_{\mu,b} \rangle + (1-f_b)
\langle \hat{S}_{\mu,g} \rangle.
\end{align}
Because of the antiferromagnetic coupling between impurity spins on the $A$ and $B$ sublattices,
in the presence of the Zeeman coupling to the spins we do not expect them to be collinear; the
mean-field state should be a canted antiferromagnet, in which $\vec{M}_A$ and
$\vec{M}_B$ have parallel components along $\vec{B}$, and antiparallel components
perpendicular to it. It is important to recognize that this is a broken symmetry
state, with the planar angle of the latter components in the groundstate determined arbitrarily;
i.e., the state has broken U(1) symmetry.
Without loss of generality, for the purposes of the mean-field solutions
we can take the magnetizations to lie in the $\hat{x}-\hat{z}$ plane, assuming
$\vec{B} = B\hat{z}$.  Note that because of the explicit Zeeman coupling, there will
always be non-zero components of $\vec{M}_A$ and $\vec{M}_B$ along the $\hat{z}$ direction
at any temperature.  The spontaneous ordering is captured by the non-zero $\hat{x}$
components of these.

\begin{figure}
  \centering
  \includegraphics[width=0.5\textwidth]{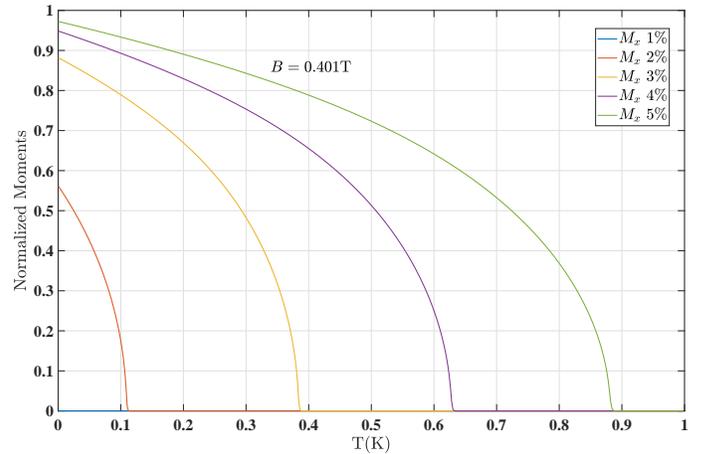}
  \caption{$\vec{M}_A\cdot\hat{x}=-\vec{M}_B\cdot\hat{x}$ evaluated in mean field theory
  for $B=0.401$T. }
  \label{MxZeeman}
\end{figure}

\begin{figure}
  \centering
  \includegraphics[width=0.5\textwidth]{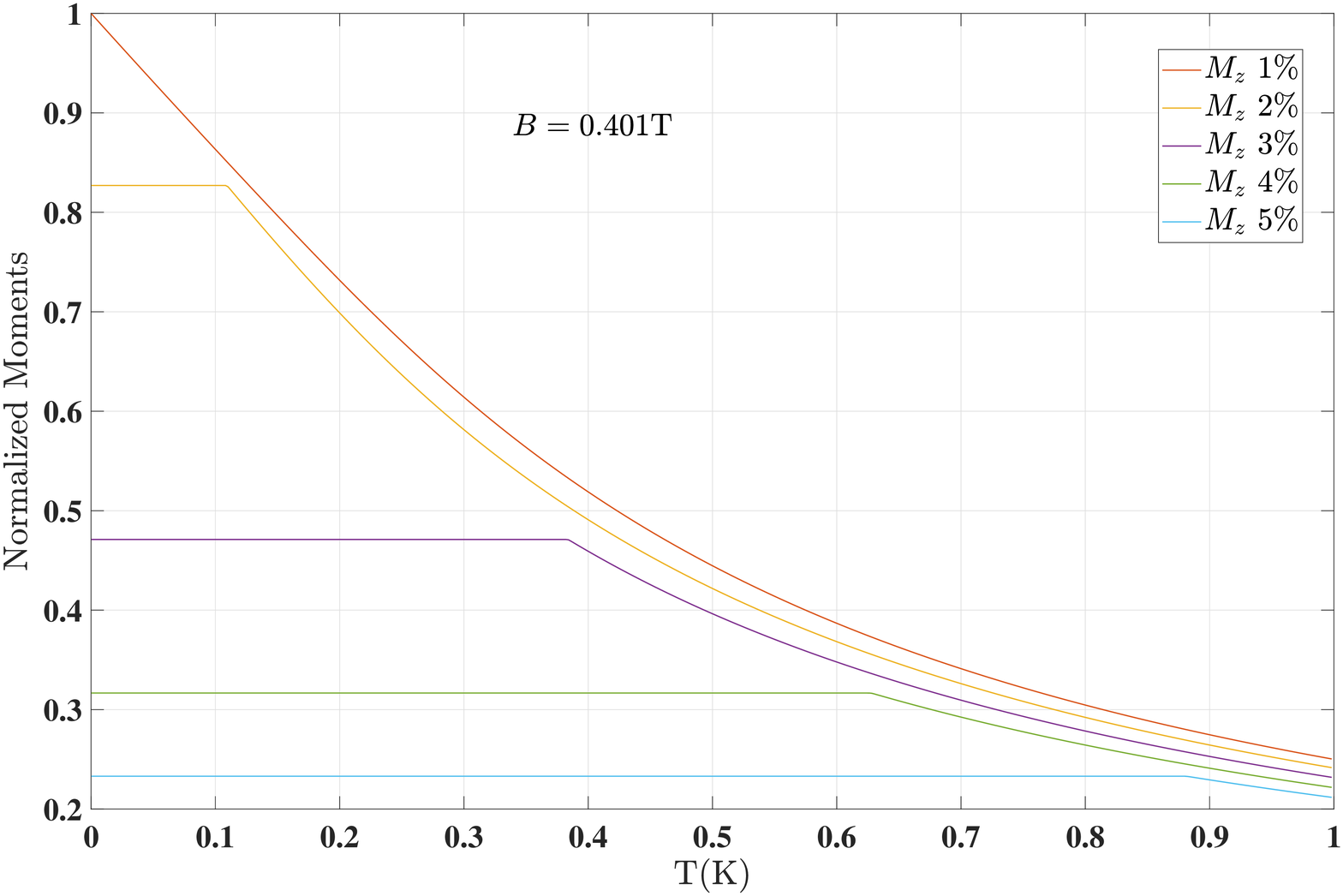}
  \caption{$\vec{M}_A\cdot\hat{z}=\vec{M}_B\cdot\hat{z}$ evaluated in mean field theory
  for $B=0.401$T. }
  \label{MzZeeman}
\end{figure}

\begin{figure}
  \centering
  \includegraphics[width=0.5\textwidth]{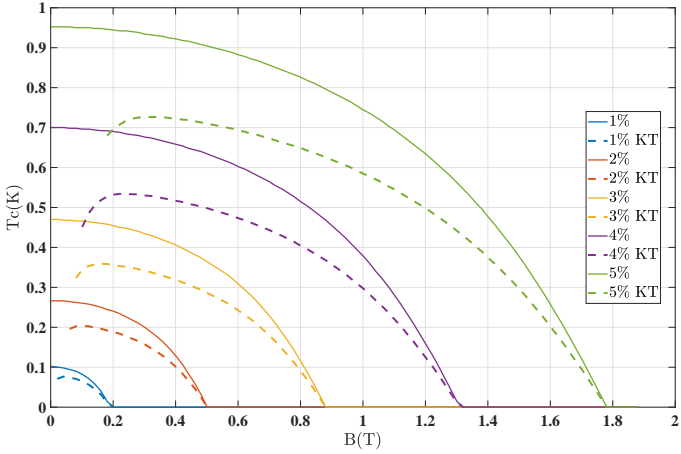}
  \caption{Mean-field phase diagrams for real magnetic field case. Schematic forms
  showing expected behavior of Kosterlitz-Thouless transitions shown as dashed lines.}
  \label{ZeemanPD}
\end{figure}

Figs. \ref{MxZeeman} and \ref{MzZeeman} illustrate typical results at $B=0.401$T.
At low temperature
$\hat{z} \cdot \vec{M}_A=\hat{z} \cdot \vec{M}_b$ becomes pinned to a value less than one at these fields, indicating that the spins have become canted.  Above a crossover temperature this
value begins to fall, showing that polar fluctuations of the spin direction begin to
become important.  By contrast, $\hat{x} \cdot \vec{M}_A=-\hat{x} \cdot \vec{M}_b$ falls
continuously with temperature, reflecting the behavior of in-plane spin fluctuations.
This component truly drops to zero at a mean-field transition temperature, and the broken
U(1) symmetry of the spin ordering is restored.  Fig. \ref{ZeemanPD} illustrates representative
mean-field phase diagrams at different magnetic fields.

There are well-known limitations to the use of mean-field theory results for systems with broken
continuous symmetries in two dimensions, such as in our analyses in this and the previous section.
We now turn to a discussion of these, as well as other implications and speculations related to
our study.

\section{Discussion and Summary}
\label{sec:sum}
In the analyses above, we considered pairwise RKKY interactions between spins in two situations where
the electronic states are organized as Landau levels, in one case generated by non-uniform strain in the system, in the other by a magnetic field.  We found that when the Fermi energy is in the
lowest Landau level (LLL), the large degeneracy of states leads to a situation in which
two pairs of states
break off from the LLL, two above and two below the Fermi energy, whose precise energies depend
on the relative orientations of the impurities.  This means that the impurity pair binds a
pair of electrons, which leads to a relatively strong interaction between them, up to a
distance of order the magnetic length.  Beyond this, the RKKY interactions are dominated by
perturbative changes to the negative energy states, which on a coarse-grained scale
leads to ferromagnetic
interactions on the same sublattice, and antiferromagnetic interactions across sublattices.
Rapid oscillations around these behaviors can be traced back to the effect of the negative
energy cutoff, which is determined by the total density of electrons in the $p_z$ orbitals
of the carbon atoms.

In practical situations, it is not sufficient to treat dilute spin impurities on the graphene
as simply coupled by the RKKY interaction formally computed above, because the density of
electrons in the LLL at field scales that are practically realizable is quite small compared
to the density of impurities.  In this situation we expect most impurities will not capture
bound electrons.  We considered the implications of this in a simple mean-field theory, in
which a fraction of impurity spins  $f_b$ binds electrons, of sufficient number to precisely
deplete the LLL.  We found that the effect of this can lead to a significant increase in
the mean-field $T_c$, but that the magnetization above $T_c$ for the situation with $f_b=0$
is very small, because $f_b$ is typically quite small.  Interestingly, in the case of
magnetic fields due to non-uniform strain, the imbalance between sublattices leads to a
stronger magnetization on one than the other at any finite temperature, so the ordered
state is formally ferrimagnetic.  Again, because of the relative smallness of $f_b$, the
net moment is small.

The nature of the broken symmetries in these two-dimensional magnets is such that we do not
truly expect spontaneous long-range magnetic order to set in at any finite temperature \cite{Mermin_1966}.
The mean-field phase diagram instead indicates a cross-over temperature at which
measurements of the spin-spin correlation length begins to exceed the typical
distance between neighboring magnetic moments.  In principle this effect could be
observed by slowly lowering the temperature of the system to some value below $T_c$,
and then rapidly quenching the system to very low temperature, so that thermal magnetic
fluctuations are frozen out.  In principle, a local measurement of the magnetization at
this low temperature would reveal domains of size the correlation length at the temperature
from which the system was quenched.

In the case of Landau levels generated by real fields, the presence of Zeeman coupling has
interesting implications.  The symmetry of the effective Hamiltonian for the spins is lowered
from SU(2) to U(1), and the mean-field groundstate when ordered is a canted antiferromagnet.
Even in the absence of long-range magnetic order, the system should still exhibit a true
thermodynamic, Kosterlitz-Thouless (KT) phase transition, in which vortices of the in-plane
component of the magnetization unbind at some temperature $T_{KT}$.  To estimate this,
we consider a very simple model, in which the U(1) spin degrees of freedom are on a
square lattice, with nearest neighbor coupling constant $J_{eff}$.  By computing the mean-field
$T_c$ of this model, and matching this to the simplest estimate of the (KT) transition,
$k_BT_{KT}=\pi J_{eff}/2$ \cite{Nelson_book}, one can obtain an estimate for $T_{KT},$ which
turns out to be simply proportional to the mean-field $T_c$, as illustrated in Fig.
\ref{ZeemanPD}.  An important caveat with respect to this estimate is that it does not
include any renormalization of $J_{eff}$ due to vortex-antivortex pairs, which always
decreases the transition temperature.  In particular this renormalization should become
quite large as $B \rightarrow 0$, and SU(2) symmetry is restored, in which case there will
be no KT transition, so that $T_{KT} \rightarrow 0$.  The low barrier to spins tilting
into the $\hat{z}$ direction in
this limit suggests that the core energy of vortices becomes small, so that our
simple estimate becomes increasingly unreliable.  Fig. \ref{ZeemanPD} schematically shows the
form we expect $T_{KT}$ to take in a more sophisticated treatment.

We conclude with some speculations about further interesting behaviors this system
might host.  In the case of non-uniform strain, as noted above, the relatively
large energy scale associated with binding of LLL electrons to sites leads to two
temperature scales, a lower one at which the majority of spins lose most of their
collective order, and higher one at which those binding electrons do so.  While direct
magnetization measurements are unlikely to detect this, an interesting possibility is
that it might be visible in transport due to scattering of electrons from these spin
degrees of freedom, or via na anomalous Hall effect.
Beyond this, the behavior of the system with greater levels
of doping may show interesting effects, for example yielding a mean-field $T_c$
that drops as the LLL is depleted.  Further doping may yield similar physics to that of
the LLL when the Fermi energy reaches negative Landau levels, which may yield oscillations
in $T_c$.  Finally, in our mean-field analysis, collective behavior among the
degrees of freedom was not considered.  For example,
correlations among nearby impurity spins could lead to
electrons binding to {\it multiple} impurities rather than individual ones.  In principle
our mean-field theory could accommodate this phenomenologically by adopting
larger values of $f_b$; since $\ell$ is typically much larger than the average
distance between impurities, this renormalization could be considerable.  If such
effects are important, we expect ferrimagnetism in a strain-induced field
could be notably larger than our estimates above.  Beyond this, interactions among the
electrons themselves may induce a spin stiffness, which could lessen the correlation
between electron and impurity spins in the system, and would tend to work against
the induced ferrimagnetism.
Thus the net effect of correlated behaviors in this system is unclear.  We leave their
consideration for future research.

\textit{Acknowledgements --}
This work was supported by the NSF through Grant Nos.
DMR-1506263 and DMR-1506460, by the US-Israel Binational Science Foundation,
and by Indiana University through an FRSP grant.
HAF thanks the Aspen Center for Physics, where
part of this work was done.

\end{document}